\documentclass[twocolumn,aps,prl,superscriptaddress,floatfix,longbibliography]{revtex4-1}  

\makeatletter
\renewcommand*{\@fnsymbol}[1]{\ensuremath{\ifcase#1\or \dagger\or *\or \ddagger\or
\mathsection\or \mathparagraph\or \|\or **\or \dagger\dagger \or
\ddagger\ddagger \else\@ctrerr\fi}} \makeatother

\usepackage{color}%

\usepackage{verbatim}
\usepackage{graphicx}
\usepackage{amssymb} 
\usepackage{dcolumn}
\usepackage{soul}
\usepackage{bm}
\usepackage[mathlines]{lineno}
\usepackage[colorlinks,linkcolor=blue,anchorcolor=blue,citecolor=blue,urlcolor=blue]{hyperref}
\usepackage{multirow}
\usepackage{color}
\usepackage{amsmath}
\usepackage{amssymb}
\usepackage[commandnameprefix=always]{changes}
\usepackage{float}

\begin{document}


\title{
LLM-driven discovery for carbon allotropes with bond-network entropy \\
}


\author{Yuzhou Hao}
\affiliation{State Key Laboratory for Mechanical Behavior of Materials, State Key Laboratory of Porous Metal Materials, School of Materials Science and Engineering, Xi'an Jiaotong University, Xi'an 710049, China}

\author{Yujie Liu}
\affiliation{State Key Laboratory for Mechanical Behavior of Materials, State Key Laboratory of Porous Metal Materials, School of Materials Science and Engineering, Xi'an Jiaotong University, Xi'an 710049, China}

\author{Xuejie Li}
\affiliation{State Key Laboratory for Mechanical Behavior of Materials, State Key Laboratory of Porous Metal Materials, School of Materials Science and Engineering, Xi'an Jiaotong University, Xi'an 710049, China}

\author{Turab Lookman}
\affiliation{AiMaterials Research LLC, Santa Fe, New Mexico 87501, USA}

\author{Xiangdong Ding}
\affiliation{State Key Laboratory for Mechanical Behavior of Materials, State Key Laboratory of Porous Metal Materials, School of Materials Science and Engineering, Xi'an Jiaotong University, Xi'an 710049, China}

\author{Jun Sun}
\affiliation{State Key Laboratory for Mechanical Behavior of Materials, State Key Laboratory of Porous Metal Materials, School of Materials Science and Engineering, Xi'an Jiaotong University, Xi'an 710049, China}

\author{Zhibin Gao}
\email{zhibin.gao@xjtu.edu.cn}
\affiliation{State Key Laboratory for Mechanical Behavior of Materials, State Key Laboratory of Porous Metal Materials, School of Materials Science and Engineering, Xi'an Jiaotong University, Xi'an 710049, China}

\date{\today}

\begin{abstract}

The discovery of novel carbon allotropes with tailored thermal and mechanical properties is critical for advanced thermal management. However, exploring the vast configurational space of carbon using \textit{ab initio} calculations remains computationally prohibitive. 
Driven by the rich topological landscape of carbon, where the competition between $sp, sp^2,$ and $sp^3$ hybridization states dictates material performance, we establish a closed-loop AI framework to explore this complex configurational space. We introduce a hybridization entropy descriptor to guide the search beyond conventional forms. 
Here, we establish a closed-loop AI framework that synergizes a Large Language Model (LLM) for structural generation with a Machine Learning Potential (MLP) for accelerated evaluation. Leveraging CrystaLLM to generate candidates and an iteratively refined MLP for high-fidelity validation, we screened thousands of structures to identify several stable allotropes with exotic properties. Specifically, we report ``yne-diamond C$_{12}$'' and ``yne-hex-diamond C$_{8}$'', which exhibit extreme thermal anisotropy and ultralow in-plane shear stiffness arising from their mixed $sp$-$sp^3$ hybridization. Furthermore, we discovered a complex $sp$-$sp^2$-$sp^3$ hybridized C$_{12}$ phase that combines metallic conductivity with an anomalous negative Poisson’s ratio. Notably, we identified a superhard phase (C16\_3) possessing a calculated Vickers hardness (103.3 GPa) exceeding that of diamond 96 GPa~\cite{ANDRIEVSKI2001447}). Microscopic analysis reveals that thermal transport in these materials is governed by the interplay between rigid frameworks and flexible linkers. 
This work expands the known carbon phase space and demonstrates the efficacy of coupling generative AI with machine learning potentials for the accelerated inverse design of functional materials.


\end{abstract}

\maketitle

Carbon materials, ranging from 3D diamond to 2D graphene and 1D carbon nanotubes~\cite{Iijima1991microtubules, Iijima1993nanotubes, Bethune1993Cobalt}, are critical to thermal management, electronics, and energy storage owing to their unique $sp$, $sp$$^2$, and $sp$$^3$ hybridization-induced properties~\cite{Balandin2008}. Despite this rich polymorphic landscape, discovering new thermodynamically stable allotropes with tailored properties (e.g., high thermal conductivity, hardness or mechanical anisotropy) remains challenging. Conventional approaches, including random structure searching and evolutionary algorithms such as USPEX~\cite{Oganov2006USPEX} and CALYPSO~\cite{Wang2010CALYPSO}, are often constrained by low search efficiency and the high computational cost of first principle calculations.


Artificial intelligence (AI) has emerged as a transformative tool in materials science. Beyond earlier generative models like Graph Neural Networks (GNNs) and Variational Autoencoders (VAEs)~\cite{Xie2018,Luo2024Deep}, Large Language Models (LLMs) such as CrystaLLM~\cite{antunes2024crystal, Song2025Inverse} have shown that treating crystallographic data as text sequences enables the generation of chemically valid structures. The scalability of such approaches was further demonstrated by the Graph Networks for Materials Exploration (GNoME)~\cite{Scaling2023Merchant}. However, a critical challenge persists: coupling LLM-based generation with a rapid, high-fidelity validation framework that can handle complex, non-equilibrium bonding environments without incurring the prohibitive computational burden of \textit{ab initio} calculations.

\begin{figure*}
\includegraphics[width=1.50\columnwidth]{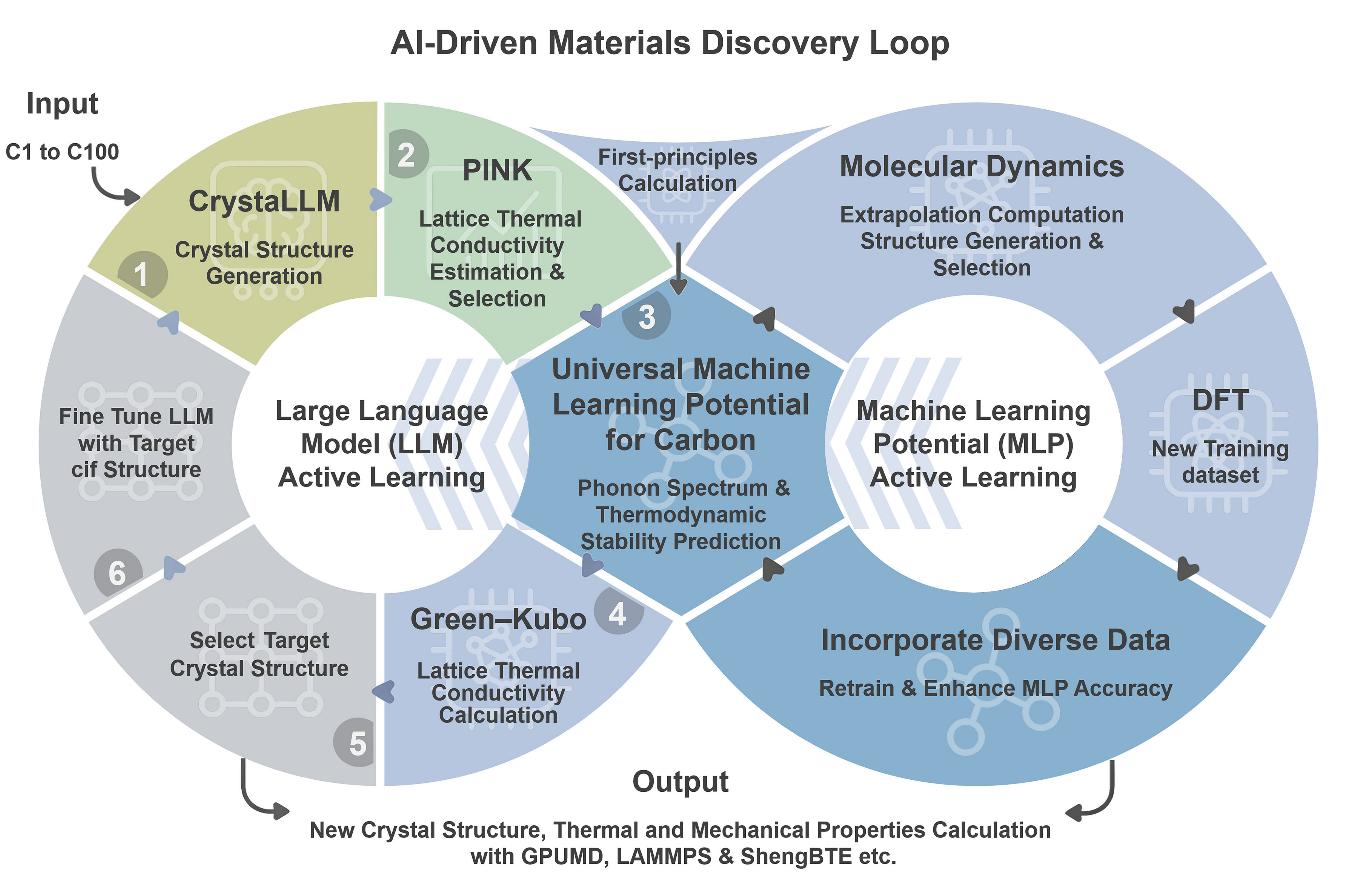}
\vspace{-2mm}
\caption{
Schematic illustration of the closed-loop AI-driven materials discovery workflow. The framework integrates two synergistic active-learning cycles. The generative cycle (left) utilizes a Large Language Model (CrystaLLM) to propose and screen candidate structures with varying atom counts (C$1$–C${100}$), followed by fine-tuning on DFT-verified stable phases. The potential training cycle (right) iteratively refines the Machine Learning Potential (MLP) using on-the-fly data generation. This combined pipeline enables rapid high-throughput screening and accurate thermal/mechanical property evaluation via GPUMD and ShengBTE simulations.
\label{fig1}}
\end{figure*}

To address this, we propose a comprehensive dual-loop active learning (AL) framework, shown in Fig.~\ref{fig1}. The first loop leverages an LLM to explore the carbon compositional space, generating candidate structures containing up to 100 atoms per conventional cell (C1–C100, optimized by MatterSim~\cite{yang2024mattersim}). These candidates are rapidly screened: first by our homegrown PINK code~\cite{Liu2025PINK} for preliminary lattice thermal conductivity ($\kappa_L$) estimation, and subsequently for dynamical stability using Phonopy~\cite{TOGO2015First} combined with the current iteration of our Machine Learning Potential (MLP). The second loop supports this generation by iteratively constructing a universal MLP, based on the Neuroevolution Potential (NEP) architecture~\cite{Xu2025GPUMD40}. Unlike standard empirical potentials, this MLP is actively retrained by sampling diverse, extreme configurations—ranging from ultra-high pressure phases ($> 500 \text{ GPa}$, shown in Fig.~\ref{fig3}(c) and strain stress relationship in Fig.~S5) to complex mixed-hybridization networks—thereby ensuring DFT-level accuracy across the entire potential energy surface. This workflow seamlessly connects generative design with high-fidelity physical property calculations (using GPUMD~\cite{Fan2022GPUMD}, LAMMPS~\cite{LAMMPS}, and CSLD~\cite{wang2023role}), culminating in the accelerated discovery of novel carbon materials with customized mechanical and thermal properties.


Here, we employ this framework to explore the configuration space of carbon, identifying stable allotropes that include high-density superhard phases and porous anisotropic frameworks.
Crucially, our search is driven by a specific physical motivation: to overcome the limitations of pure $sp^3$ networks (which are thermally isotropic and brittle) by exploring the 'mixed-hybridization' topological space. We hypothesize that introducing $sp$-hybridized acetylenic linkers or $sp^2$ nodes into a rigid $sp^3$ framework can decouple thermal transport from mechanical stiffness. This ``phonon engineering'' strategy aims to identify metastable phases where the interplay between rigid frameworks and flexible linkers enables exotic functionalities. 
We focus on three exemplary structures: yne-diamond C$_{12}$, yne-hex-diamond C$_{8}$, and a complex $sp$-$sp^2$-$sp^3$ hybrid C$_{12}$~\cite{Matar2024Novel,Samir2023ene,Meng2014Superhard,ZHANG2020109904}. Combining the Boltzmann Transport Equation (BTE) with heterogeneous non-equilibrium molecular dynamics (HNEMD)~\cite{Gabourie2021Spectral}, we elucidate the microscopic mechanisms governing thermal transport in these systems. These results demonstrate the critical role of AI-assisted design in discovering materials with extreme properties beyond the reach of traditional screening methods.

\begin{figure*}
\includegraphics[width=2.0\columnwidth]{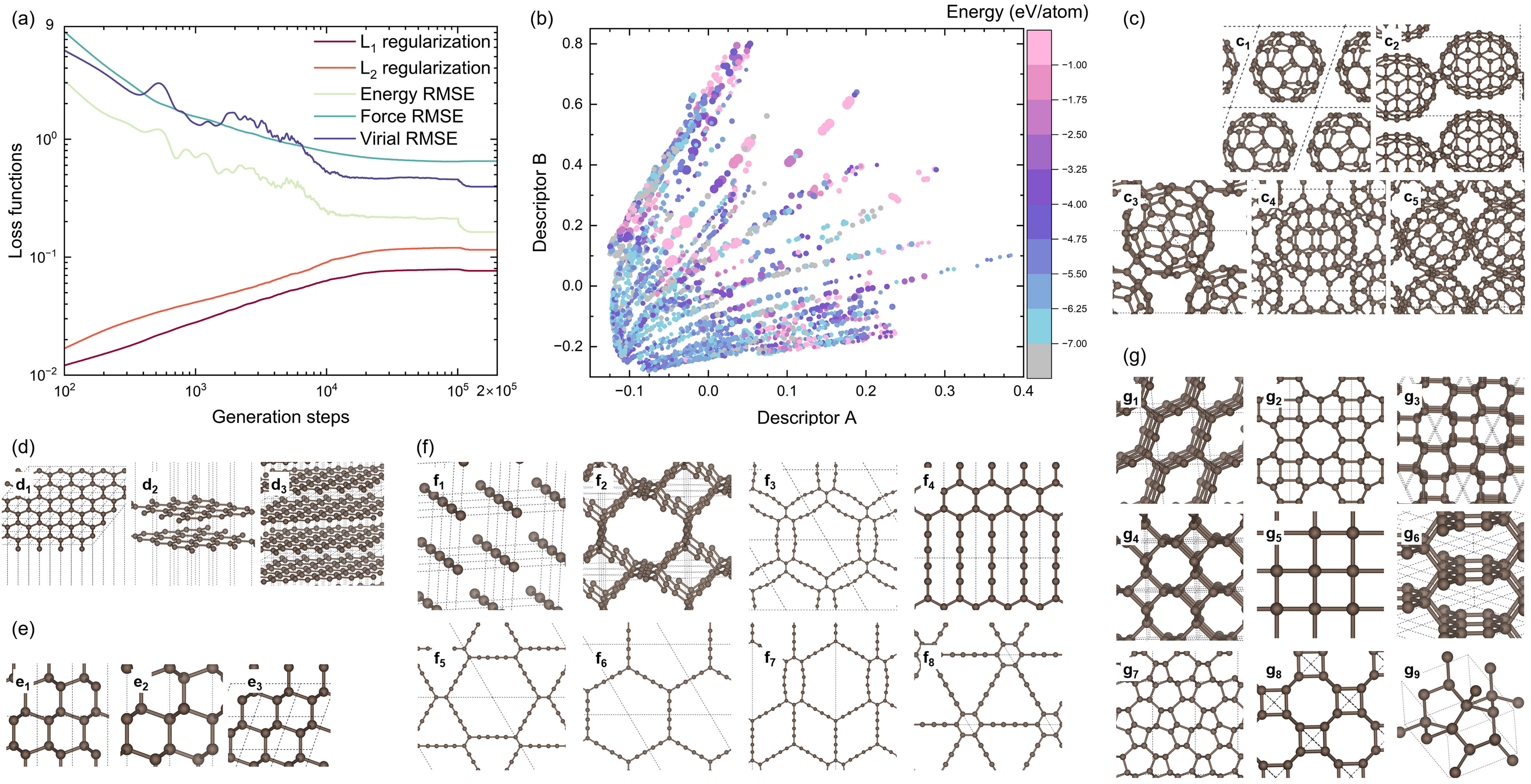}
\vspace{-2mm}
\caption{
Training performance and dataset diversity of the carbon Machine Learning Potential (NEP). (a) Evolution of loss functions (including L$_1$/L$_2$ regularization) and RMSE values for energy, force, and virial versus training steps, illustrating model convergence. (b) Distribution of atomic environments in the descriptor space, color-coded by energy per atom, demonstrating comprehensive structural coverage. (c)–(g) Representative structures from the training set, including fullerenes (c$_1$-c$_6$), layered graphene (d$_1$-d$_3$), diamond (e$_1$-e$_3$), $sp$ hybridization states carbon (f$_1$-f$_8$) and diverse 3D frameworks (g$_1$-g$_9$).
\label{fig2}}
\end{figure*}

\begin{figure*}
\includegraphics[width=1.5\columnwidth]{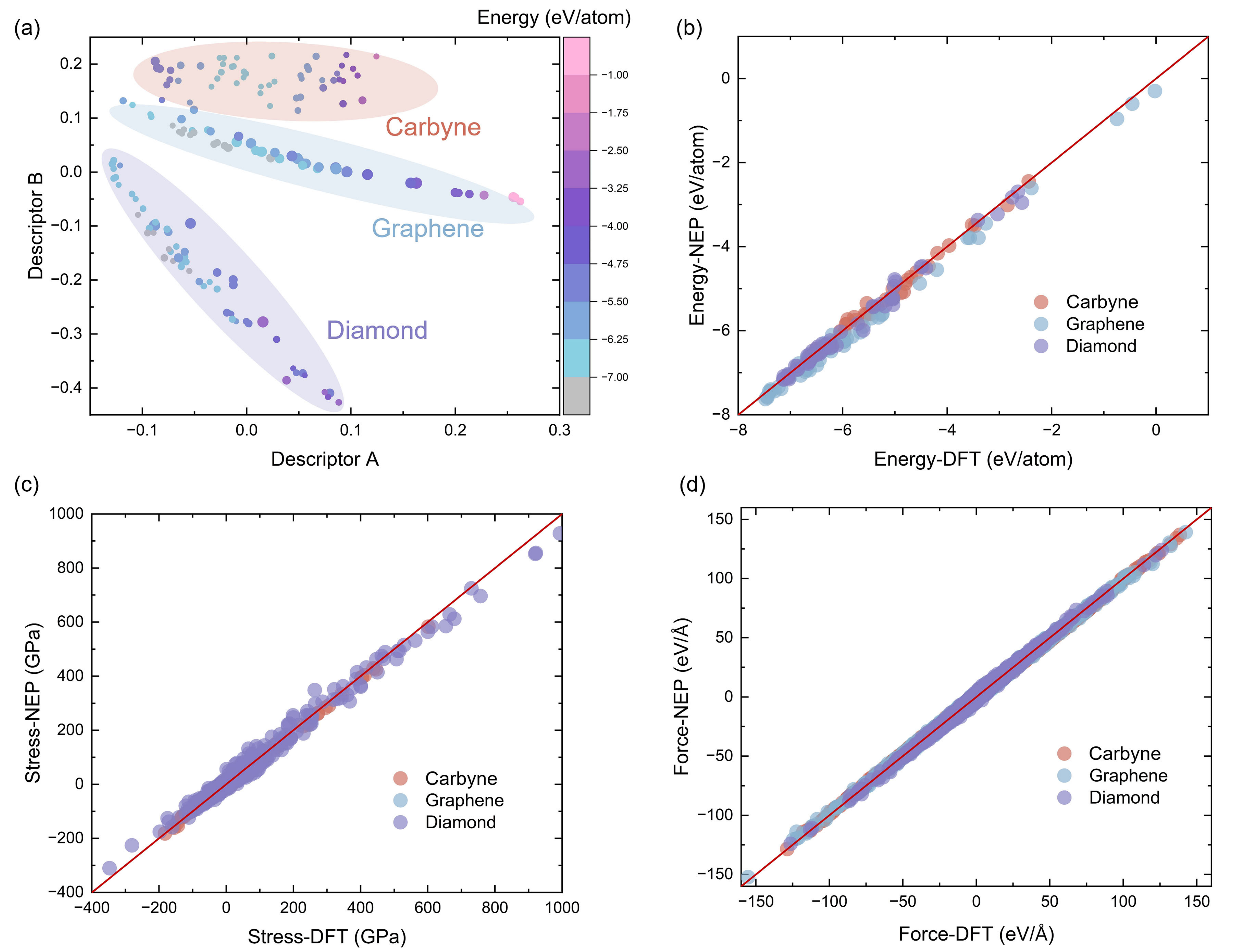}
\vspace{-2mm}
\caption{
Validation of the NEP model against DFT benchmarks. (a) Visualization of atomic environments in the descriptor space, showing distinct clustering for carbyne (red), graphene (light blue), and diamond (dark blue). (b)–(d) Parity plots comparing NEP predictions with DFT calculations for (b) energy per atom, (c) virial stress, and (d) atomic forces. The diagonal lines indicate perfect agreement.
\label{fig3}}
\end{figure*}

The NEP training dataset covers a comprehensive range of structural configurations and stress states. As shown in Fig.~\ref{fig2}(a), the loss functions exhibit excellent convergence: over $2 \times 10^5$ steps, the root-mean-square error (RMSE) for energy and force decreases to 0.16 eV/atom and 0.65 eV/\AA, respectively, with the virial RMSE reaching 0.40 eV/atom. These trends suggest that sufficient accuracy for phonon calculations is achieved beyond $10^5$ steps. Additionally, stable L$_1$ and L$_2$ regularization curves indicate effective complexity control, ensuring model transferability across different $sp$-hybridized structures without overfitting.


Fig.~\ref{fig2}(b) illustrates the atomic environment distribution within the descriptor space, with Descriptors A and B ranging from -0.1 to 0.4 and -0.2 to 0.8, respectively. Data points are color-mapped by energy [-7.0 eV/atom (dark blue) to -1.0 eV/atom (pink)], with marker sizes scaled to represent the energy fitting error. This extensive coverage verifies the training set's capability to represent a wide spectrum of bonding and stress configurations~\cite{CHEN2025109859}.


Fig.~\ref{fig2}(c) displays representative fullerene derivatives (c$_1$-c$_5$), ranging from isolated $C_{60}$ to 1D $C_{70}$ chains and monolayer fullerene networks (c$_3$). Layered graphene configurations (d$_1$-d$_3$) with distinct stacking sequences and vacancy defects are shown in Fig.~\ref{fig2}(d), providing insight into anisotropic thermal transport. Fig.~\ref{fig2}(e) highlights diamond polytypes (e$_1$-e$_3$), specifically comparing hcp-diamond (e$_1$) and fcc-diamond (e$_3$) phases. Furthermore, Fig.~\ref{fig2}(f) presents $sp$-hybridized allotropes (f$_1$-f$_8$), such as carbyne (f$_1$) and graphdiyne (f$_8$), which have gained prominence for their exceptional properties since their theoretical prediction~\cite{Li2010Architecture,Long2011Electronic}.


Fig.~\ref{fig2}(g) further illustrates complex 3D network structures (g$_1$-g$_9$). By encompassing the full range of hybridization states ($sp$, $sp^2$, $sp^3$) and dimensionalities (1D–3D), the dataset ensures the NEP’s applicability to diverse carbon materials for thermal and mechanical properties. Crucially, the active learning workflow systematically targets high-uncertainty structural configurations, thereby optimizing the training efficiency.

To evaluate the transferability of the carbon NEP, we validate it against an independent test set containing diamond ($sp^3$), graphene ($sp^2$), and carbyne ($sp$). As shown in Fig.~\ref{fig3}(a), the allotropes form distinct clusters in the descriptor space: diamond (purple), graphene (blue), and carbyne (red). This separation demonstrates the model's ability to discriminate between local environments characteristic of different hybridization states. In the plot, color denotes atomic energy, and marker size corresponds to the prediction error relative to density functional theory (DFT) data. Fig.~\ref{fig3}(b-d) compare NEP predictions against DFT benchmarks calculated using the optB86b functional (to account for van der Waals interactions). Fig.~\ref{fig3}(b) highlights the energy accuracy, showing exceptional linearity over a wide energy window. The model achieves high fidelity in both low-energy equilibrium states and high-energy distorted configurations, a prerequisite for robust thermodynamic stability analysis.



The robustness of the potential under extreme mechanical conditions is further evidenced by the virial stress and atomic force predictions. Fig.~\ref{fig3}(c) shows that the predicted virial stresses align with DFT data over a remarkably broad range, from tensile states of $ -400$ GPa to compression exceeding $1000$ GPa. This dynamic range, rarely achieved by empirical potentials, ensures the description of severe strain without non-physical artifacts. Similarly, Fig.~\ref{fig3}(d) demonstrates high accuracy in atomic force predictions up to $150$ eV/\AA. Capturing such extreme forces and stresses is critical for ensuring MD simulation stability, enabling the reliable exploration of phase transitions and mechanical properties under extreme pressures.

\begin{figure*}
\includegraphics[width=2.0\columnwidth]{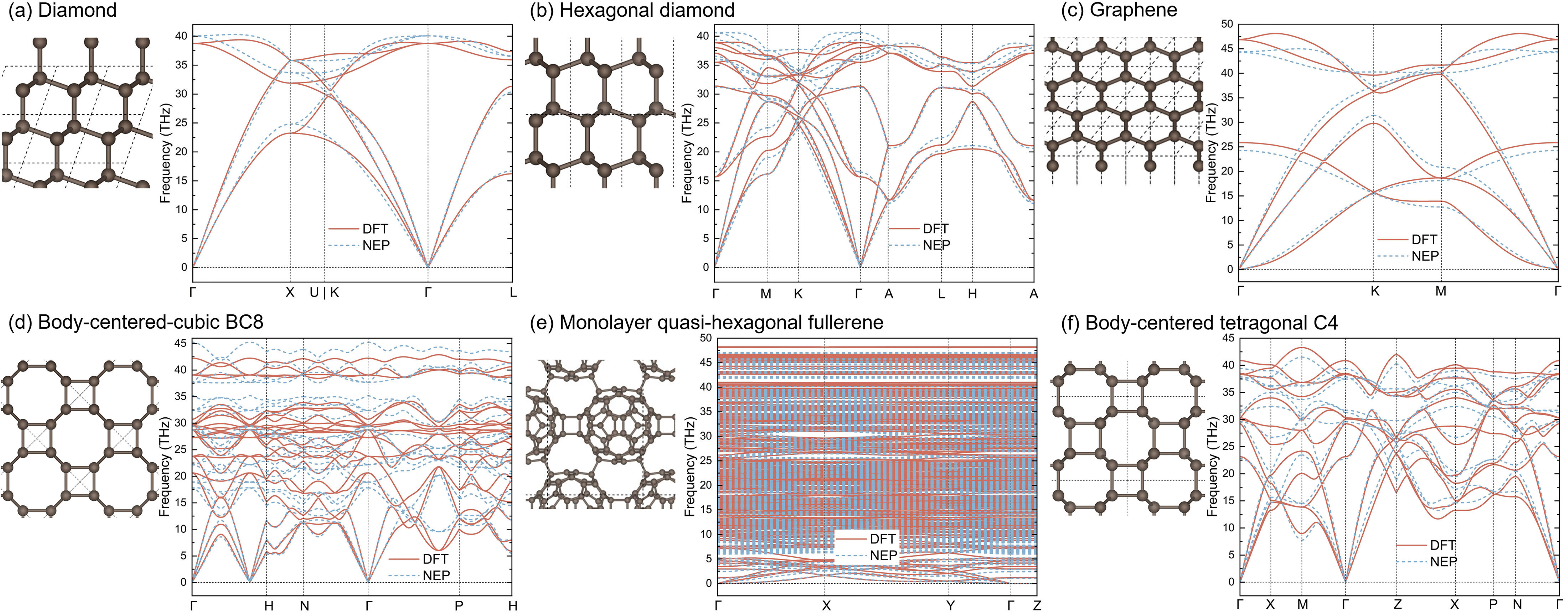}
\vspace{-2mm}
\caption{
Phonon dispersion relations for representative carbon allotropes. Comparison between predictions from the NEP model (dashed blue lines) and DFT benchmarks (solid red lines). The panels show: (a) cubic diamond, (b) hexagonal diamond, (c) monolayer graphene, (d) body-centered cubic BC8, (e) monolayer quasi-hexagonal fullerene, and (f) body-centered tetragonal C$_4$.
\label{fig4}}
\end{figure*}


We calculated phonon dispersion relations for six representative carbon allotropes (Fig.~\ref{fig4}(a–f)) to verify the NEP's capability in capturing lattice dynamics across distinct dimensionalities and hybridization. As rigorous benchmarks, Figs.~\ref{fig4}(a–c) show the phonon spectra for ideal $sp^3$ networks (cubic and hexagonal diamond~\cite{Zhang2025Thermal,Yang2025Synthesis}) and the archetypal 2D $sp^2$ system (graphene). To further test transferability to metastable phases relevant to high-pressure synthesis and generative design, we analyzed the body-centered cubic BC8 phase (Fig.~\ref{fig4}(d))~\cite{Ge2025Crystalline, Zhang2017Hydrothermal, Liu2008Synthesis, Johnston1989Superdense}, the monolayer quasi-hexagonal fullerene network (Fig.~\ref{fig4}(e))~\cite{DONG2023123943, Peng2025Monolayer, Peng2022Monolayer, Hou2022Synthesis}, and the body-centered tetragonal C4 allotrope (Fig.~\ref{fig4}(f))~\cite{Umemoto2010Body,LI2025116070}.


The NEP predictions (dashed blue lines) show remarkable agreement with DFT benchmarks (solid red lines) throughout the Brillouin zone. The acoustic branches, critical for thermal transport, are reproduced with high fidelity, especially near the $\Gamma$ point (determining group velocities). Similarly, the excellent alignment of optical modes confirms the accurate description of C-C bond stiffness across varying $sp^2$ and $sp^3$ environments. Finally, the absence of imaginary modes in all calculations verifies that the NEP correctly identifies these diverse allotropes as dynamically stable energy minima.


Given the model's accuracy in the harmonic regime, we further, using the homogeneous non-equilibrium molecular dynamics (HNEMD) method, validate its prediction of lattice thermal conductivity ($\kappa_L$) which is very sensitive to anharmonic scattering. As summarized in Table S1 (Supplementary Material), the NEP correctly reproduces the established thermal transport hierarchy. Specifically, compared to the high-$\kappa_L$ diamond and graphene, the complex metastable phases (BC8, C4, and quasi-hexagonal fullerene) exhibit significantly reduced $\kappa_L$ due to enhanced phonon scattering arising from lower symmetry and complex crystal structures. This quantitative consistency confirms the NEP's reliability for high-throughput thermal and mechanical property screening.

\begin{figure*}
\includegraphics[width=2.0\columnwidth]{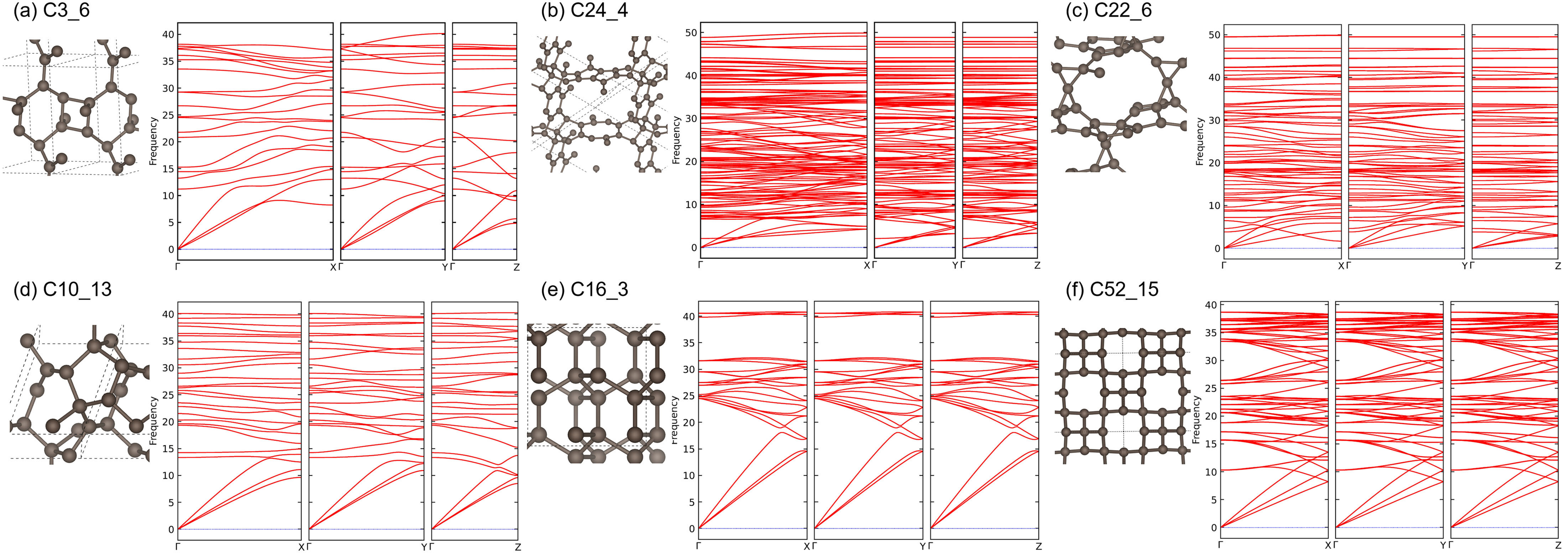}
\vspace{-2mm}
\caption{
Newly discovered stable carbon allotropes and their phonon spectra. (a)–(f) Representative structures generated by the LLM alongside their phonon dispersion relations calculated by the NEP. The structures are labeled as: (a) C3\_6, (b) C24\_4, (c) C22\_6, (d) C10\_13, (e) C16\_3, and (f) C52\_15. Here, the notation C$n\_{m}$ denotes the $m$-th predicted candidate from a generative prompt for $n$ atoms in the conventional cell. The absence of imaginary frequencies confirms the dynamical stability of all shown phases.
\label{fig5}}
\end{figure*}


Utilizing the AI-driven workflow, we screened thousands of CrystaLLM-generated candidates to isolate dynamically stable carbon allotropes. Fig.~\ref{fig5} displays six representative phases (e.g., C52\_15, denoting the 15th candidate with a target size of 52 atoms/cell) alongside their NEP-calculated phonon spectra. Crucially, all six structures exhibit real frequencies throughout the Brillouin zone, confirming their status as dynamically stable local minima on the potential energy surface. This high hit rate underscores the efficacy of our active learning strategy. Detailed statistics on candidates discarded due to thermodynamic or dynamic instability are provided in the Supplementary Material to demonstrate the rigor of the filtering process.


These allotropes display a broad spectrum of topologies, spanning open frameworks to dense networks, which drives their distinct physical behaviors. A prime example is C22\_6 (Fig.~\ref{fig5}(c)), distinguished by its porous, ring-based architecture. This structural anisotropy engenders highly directional chemical bonding and, consequently, extreme anisotropy in lattice thermal conductivity, offering significant potential for directional heat channeling. In contrast, the C16\_3 phase (Fig.~\ref{fig5}(e)) features a dense $sp^3$-dominant network, yielding a potential superhard material. Its calculated hardness reaches 103.3 GPa calculated via the Chen model~\cite{CHEN20111275}, exceeding the value for diamond obtained in this 
framework (91.8 GPa) as well as previous theoretical 93.6 GPa~\cite{Gao2003Hardness} and experimental 96 GPa~\cite{ANDRIEVSKI2001447}). In addition, structures like C24\_4 (Fig.~\ref{fig5}(b)) and C52\_15 (Fig.~\ref{fig5}(f)) exhibit complex lattices with mixed hybridization, providing a tunable trade-off between porosity and mechanical stability.



We further evaluated the mechanical and thermal performance of the discovered phases. As summarized in Table S1, we computed $\kappa_L$, density, and Vickers hardness (calculated via the Chen model~\cite{CHEN20111275}: $H_v = 2\left(k^2 G\right)^{0.585} - 3$). The results span a broad spectrum of $\kappa_L$ values, governed by the specific structural complexity and lattice anisotropy of each allotrope. The successful identification of extreme phases—specifically the superhard C16\_3 and the highly anisotropic C22\_6 demonstrates the generative model's unique ability to navigate ``hidden'' regions of the structural landscape, surpassing the limitations of conventional heuristic searches.

\begin{figure*}
\includegraphics[width=2.0\columnwidth]{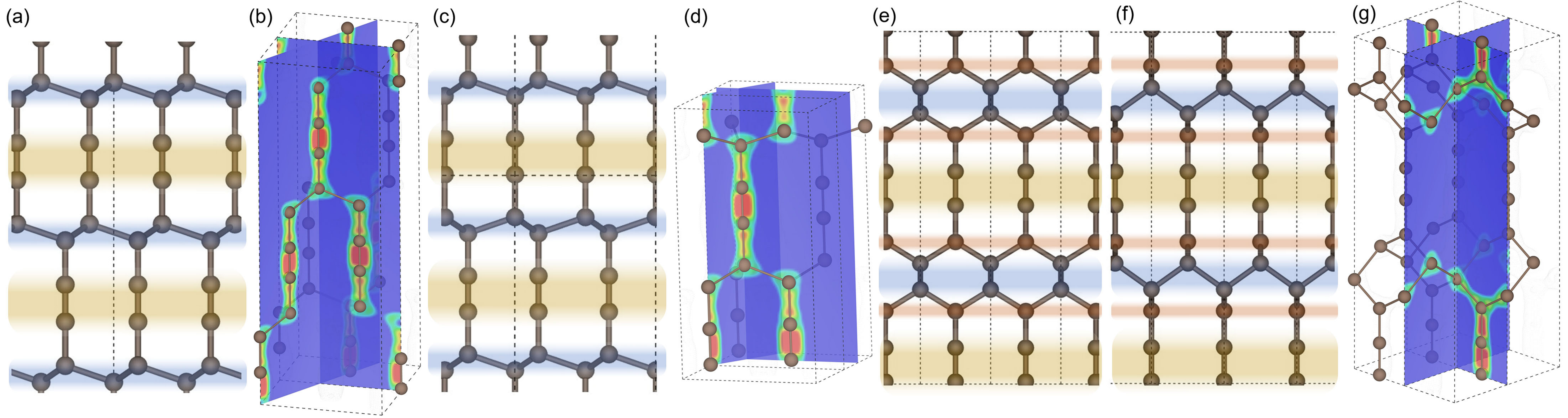}
\vspace{-2mm}
\caption{
Structural and charge density distributions of standout allotropes. (a, b) yne-diamond C$_{12}$; (c, d) yne-hex-diamond C$_{8}$; and (e–g) the $sp$-$sp^2$-$sp^3$ hybridized C$_{12}$ [(e) front view, (f) side view, (g) charge density]. In the structural models, atoms are color-coded by hybridization: $sp^3$ (blue), $sp^2$ (orange), and $sp$ (yellow). The corresponding charge density maps [(b), (d), (g)] illustrate electron distribution, with red/yellow isosurfaces indicating accumulation and blue regions indicating depletion.
\label{fig6}}
\end{figure*}


Among the discovered phases, three standout structures exhibit unique mixed-hybridization topologies (Fig.~\ref{fig6}). We term the first two ``yne-diamond $C_{12}$'' (Fig.~\ref{fig6}(a)) and ``yne-hex-diamond $C_{8}$'' (Fig.~\ref{fig6}(c))~\cite{Matar2024Novel, Samir2023ene, Meng2014Superhard}. Analogous to graphdiyne, these structures are conceptually formed by inserting linear $sp$-chains (acetylenic linkages) into the C-C bonds of cubic and hexagonal diamond, respectively. The hybridization hierarchy is visualized in Fig.~\ref{fig6} (blue: $sp^3$, yellow: $sp$, orange: $sp^2$). Furthermore, the charge density maps (Fig.~\ref{fig6}(b, d)) show strong electron localization along the bond axes, indicative of insulating or semiconducting behavior.

\begin{figure*}
\includegraphics[width=2.0\columnwidth]{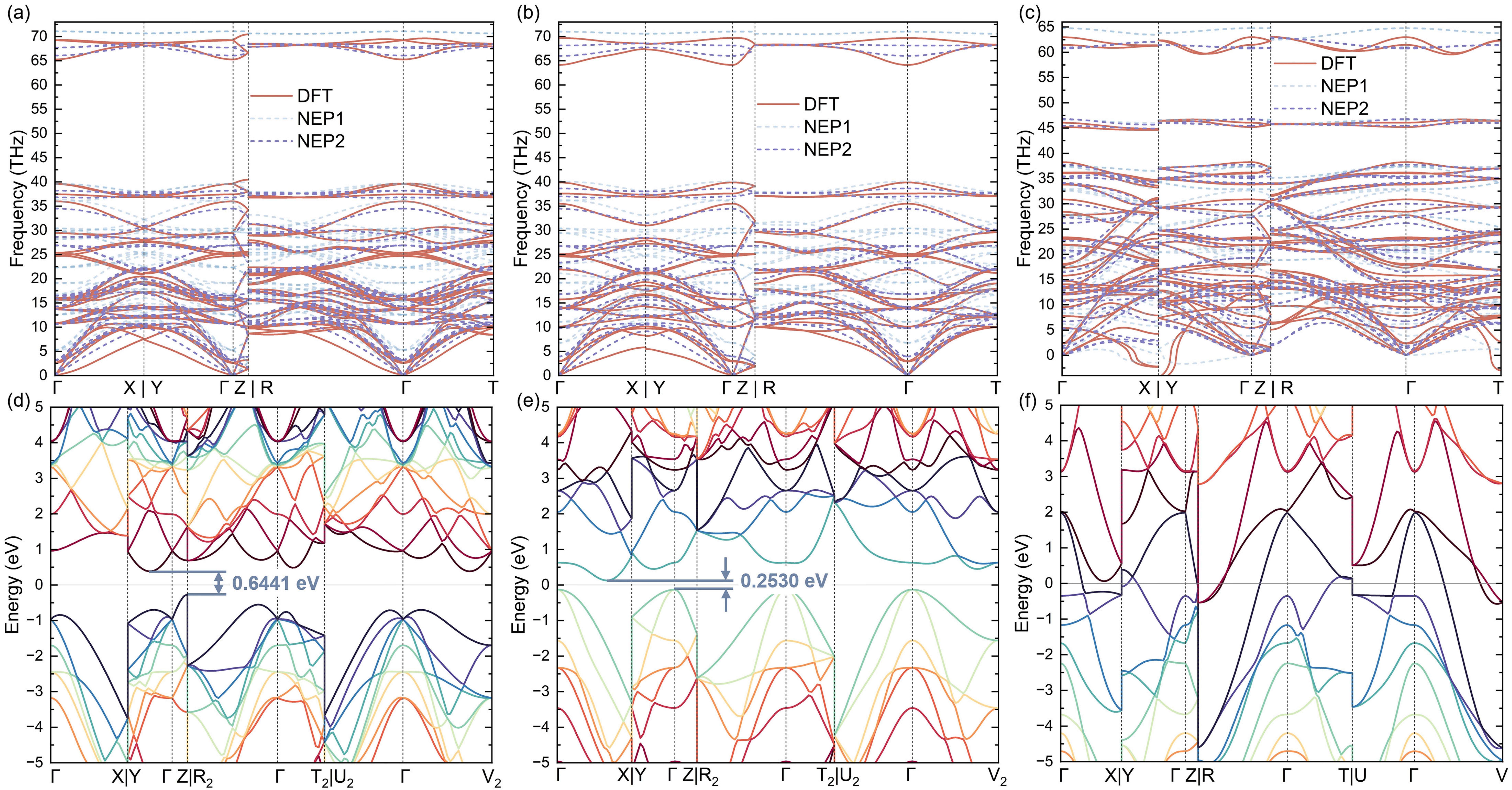}
\vspace{-2mm}
\caption{
Comparison of phonon dispersion and electronic band structures for the three standout allotropes. (a)–(c) Phonon spectra for yne-diamond C$_{12}$, yne-hex-diamond C$_{8}$, and the $sp$-$sp^2$-$sp^3$ hybridized C$_{12}$. The plots compare predictions from the initial universal potential (NEP1, dashed light blue lines) and the refined potential (NEP2, dashed purple lines) against ground-truth DFT data (solid red lines), highlighting improvements from active learning iterations. (d)–(f) Corresponding electronic band structures for the same three phases.
\label{fig7}}
\end{figure*}

The third structure, an $sp$-$sp^2$-$sp^3$ hybridized C$_{12}$ (Figs.~\ref{fig6}(e, f)), features a complex ternary hybridization containing all three carbon coordination states~\cite{ZHANG2020109904}. Its architecture displays a unique layered motif: rigid $sp^3$ frameworks bridged by $sp$-$sp^2$ networks. Crucially, the charge density map (Fig.~\ref{fig6}(g)) indicates significant electron delocalization along these $sp$-$sp^2$ linkages, pointing to a metallic nature. This is corroborated by subsequent band structure calculations (Fig.~\ref{fig7}), which classify this C$_{12}$ phase as an electrical conductor, distinct from the insulating behavior typical of pure $sp^3$ networks.


To comprehensively assess the performance potential of these three allotropes, we calculated their mechanical and thermal properties, as summarized in Table S1 (Supplementary Material). The data highlight remarkable mechanical anisotropy. Specifically, while these structures maintain a high Young's modulus along the covalent backbone (indicating high axial stiffness), they exhibit an ultralow in-plane shear modulus that is a feature often associated with lubricity or flexibility in specific directions. Most notably, the $sp$-$sp^2$-$sp^3$ hybridized structure exhibits a negative Poisson's ratio. This rare auxetic behavior, where the material expands laterally under tension, is highly unusual for carbon networks~\cite{Gao2017Novel,ZHANG2020109904}. However, it is important to note that this specific structure approaches the limit of dynamical stability. Phonon calculations reveal soft imaginary phonon modes near the $\Gamma$ point, and \textit{ab initio} molecular dynamics (AIMD) simulations suggest dynamical stability is maintained only up to approximately 100 K. Despite this low-temperature limitation, the unique coexistence of metallic conductivity and auxeticity offers valuable insights into the design rules for exotic carbon functional materials.

To rigorously verify prediction reliability and elucidate the electronic nature of these discovered materials, we calculated their phonon dispersion relations and electronic band structures, as presented in Fig.~\ref{fig7}. Figs.~\ref{fig7}(a–c) display the phonon spectra for yne-diamond C$_{12}$, yne-hex-diamond C$_{8}$, and the $sp$-$sp^2$-$sp^3$ hybrid C$_{12}$, respectively. To explicitly demonstrate the efficacy of the active learning strategy (see Fig.~\ref{fig1}), we compare the results from two generations of the potential: the initial universal potential (NEP1, dashed light blue lines) and the final, iteratively refined potential (NEP2, dashed purple lines), plotted alongside the benchmark DFT data (solid red lines). Although NEP1 reproduces the dispersion, it deviates significantly in the high-frequency optical branches governing bond stretching. Conversely, the refined NEP2 achieves near-perfect agreement with DFT throughout the frequency spectrum. This enhancement underscores the efficacy of the active learning strategy in identifying and learning from high-uncertainty configurations. Consequently, the first two structures show no imaginary modes, confirming dynamical stability. For the third phase (Fig.~\ref{fig7}(c)), soft modes near $\Gamma$ indicate lattice flexibility, consistent with the metastability observed in AIMD.

In addition to stability, these allotropes offer tunable electronic properties. Using the HSE06 hybrid functional for accurate bandgap assessment, we find that yne-diamond C$_{12}$ (Fig.~\ref{fig7}(d)) and yne-hex-diamond C$_{8}$ (Fig.~\ref{fig7}(e)) are indirect semiconductors with narrow bandgaps of 0.64 eV and 0.25 eV, respectively, which are ideal for infrared or thermoelectric applications. Conversely, the $sp$-$sp^2$-$sp^3$ hybrid (Fig.~\ref{fig7}(f)) is metallic, with bands crossing the Fermi level. As indicated by the charge density (Fig.~\ref{fig6}), this conductivity arises from $\pi$-electron delocalization in the $sp$-$sp^2$ network. This demonstration of tuning carbon from a wide-bandgap insulator to semiconductors and metals via hybridization topology underscores the power of our generative framework.

\begin{figure*}
\includegraphics[width=2.0\columnwidth]{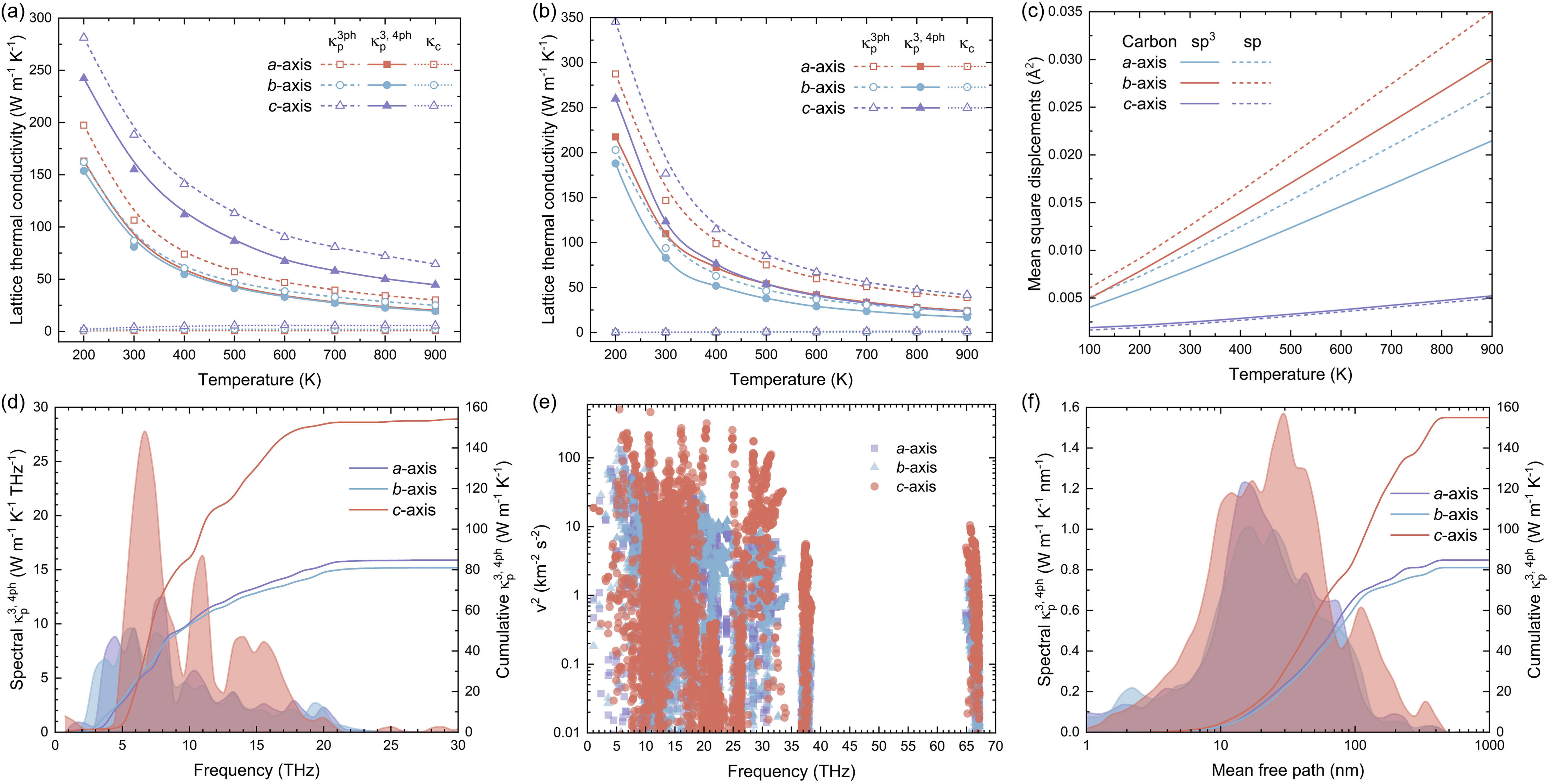}
\vspace{-2mm}
\caption{
Microscopic analysis of thermal transport mechanisms. (a, b) Temperature-dependent lattice thermal conductivity ($\kappa_L$) for yne-diamond C$_{12}$ and yne-hex-diamond C$_{8}$, decomposed into particle-like ($\kappa_p$) and wave-like ($\kappa_c$) contributions. (c) Directional mean-square displacement (MSD) of yne-diamond C$_{12}$. (d)–(f) Spectral analysis for yne-diamond C$_{12}$ at 300 K, considering both three-phonon (3ph) and four-phonon (4ph) scattering: (d) spectral and cumulative $\kappa_L$ vs. frequency; (e) phonon group velocity; and (f) spectral and cumulative $\kappa_L$ as a function of the mean free path.
\label{fig8}}
\end{figure*}
To elucidate the microscopic thermal transport mechanisms in these newly discovered allotropes, we performed a detailed analysis of their $\kappa_L$. Fig.~\ref{fig8} presents the temperature-dependent thermal properties for yne-diamond C$_{12}$ and yne-hex-diamond C$_{8}$. Specifically, Figs.~\ref{fig8}(a, b) display the calculated $\kappa_L$ along the principal crystallographic axes from 200 to 900 K. Utilizing the Boltzmann Transport framework, we decompose $\kappa_L$ into particle-like propagation ($\kappa_p$)~\cite{WuLI2014ShengBTE, Han2022FourPhonon} and wave-like coherence ($\kappa_c$) contributions~\cite{Simoncelli2019}. The results indicate that $\kappa_L$ is overwhelmingly dominated by $\kappa_p$, while $\kappa_c$ which is raised from complex inter-band interference remains negligible across the entire temperature range. This dominance confirms that, despite the complex hybridization, heat transport is governed by well-defined phonon wave packets consistent with a crystalline phonon gas model. Furthermore, we assessed the impact of higher-order anharmonicity by comparing results from three-phonon scattering only ($\kappa_{p}^{3ph}$) against those including four-phonon processes ($\kappa_{p}^{3,4ph}$)~\cite{feng2024relation, Xia2020Throughput}. The overlapping curves suggest that four-phonon scattering plays a minor role along the stiff $a$- and $b$-axes. However, along the $c$-axis, higher-order anharmonicity becomes significant, leading to a noticeable reduction in thermal conductivity for both materials~\cite{Hao2024Machine, Wang2024Anomalous, Wang2024thermoelectricity, Wang2024Revisiting, FENG2024anharmonicity, Hao2025Copper}. As shown in Fig. S6(b), increasing ngrid by 1 from its current value (8$\times$5$\times$3 for 3, 4ph processes and 9$\times$6$\times$3 for 3ph processes in yne-diamond C$_{12}$) still yields a nearly 10\% rise in certain directions, while the current calculations are approaching the limits of existing computational hardware and tools. We look forward to more accurate converged results in the future. Currently, the high-order anharmonicity and BTE calculations of unit cells with a large number of atoms remain an open question.


Both allotropes exhibit strong thermal anisotropy (Figs.~\ref{fig8}(a, b)). To identify the atomistic drivers, we analyzed the mean square displacement (MSD) in C$_{12}$ (Fig.~\ref{fig8}(c)). The results show that atomic vibrations along the $a$ and $b$ axes significantly exceed those along the $c$-axis. This finding aligns with the transport data, confirming that the rigid $c$-axis supports superior heat conduction. Species-resolved MSD analysis further highlights that the $sp$-hybridized acetylenic linkers possess much larger vibrational amplitudes than the $sp^3$ atoms. These large-amplitude vibrations of the $sp$ chains in the $a-b$ plane enhance phonon scattering, thereby lowering in-plane $\kappa_L$. In contrast, the rigid $sp^3$ backbone along the $c$-axis minimizes atomic displacement, ensuring high axial thermal transport.

We further elucidated the heat carrier spectrum by analyzing yne-diamond C$_{12}$ (Fig.~\ref{fig8}(d–f)). As shown in Fig.~\ref{fig8}(d), heat transport is dominated by acoustic and low-frequency optical phonons ($\textless$ 15 THz). These modes exhibit distinct spectral peaks corresponding to high group velocities, as confirmed by the $v^2$ distribution in Fig.~\ref{fig8}(e). Crucially, the cumulative thermal conductivity resolved by mean free path (MFP) in Fig.~\ref{fig8}(f) reveals that the primary heat carriers possess MFPs between 10 nm and 100 nm. This finding implies that nanostructuring or grain boundary engineering targeting this length scale would be an effective strategy for tuning the thermal properties of these allotropes. Collectively, our results establish yne-diamond $C_{12}$ as a thermodynamically stable phase with pronounced thermal anisotropy. This transport characteristic originates from the structural contrast between the rigid $sp^3$ backbone and the soft $sp$ linkers, which act as phonon scattering centers.

\begin{figure}
\includegraphics[width=1.0\columnwidth]{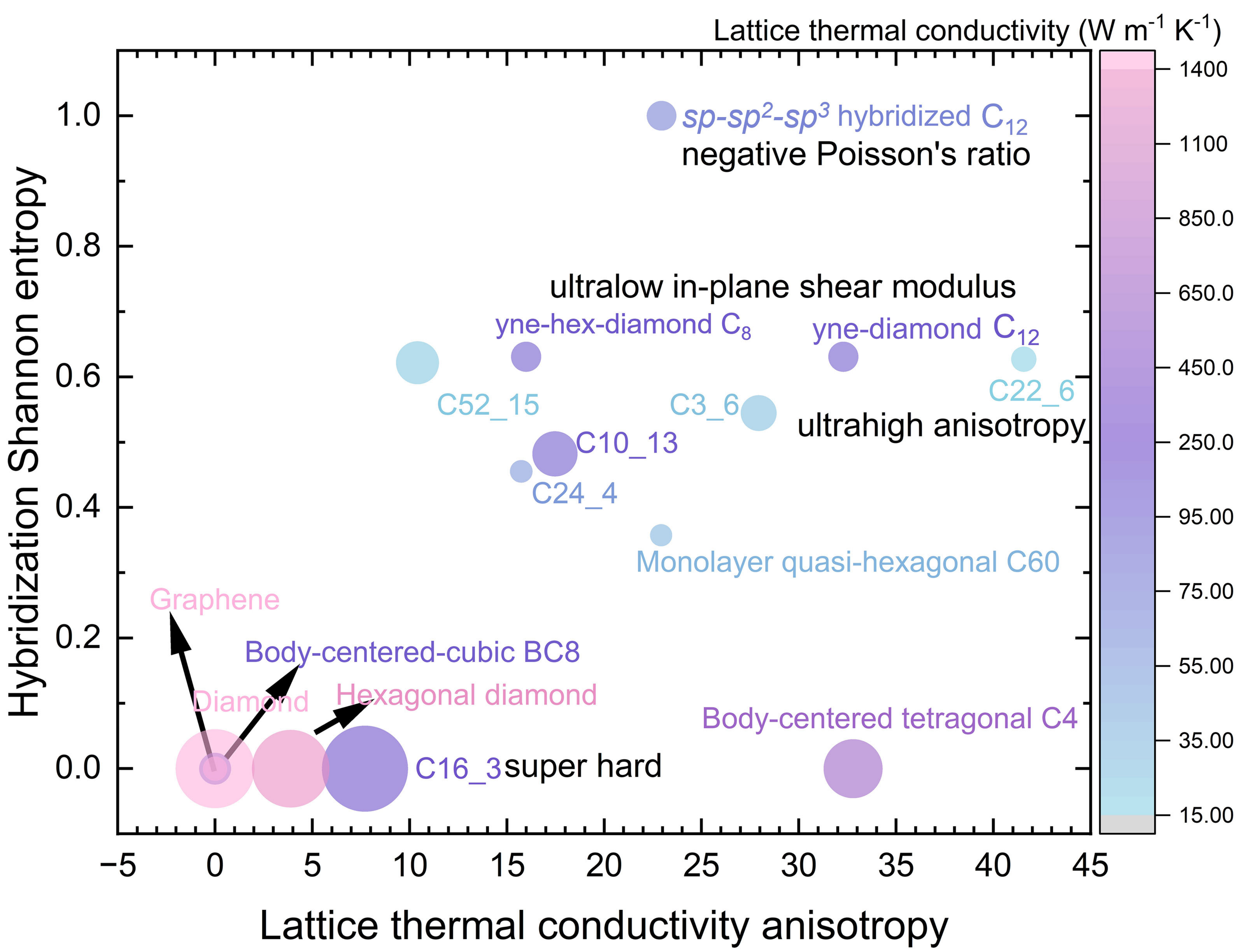}
\vspace{-2mm}
\caption{
Multi-dimensional property landscape of discovered carbon allotropes versus established phases (Eq.~\eqref{eq1} and~\eqref{eq2}). The horizontal axis represents lattice thermal conductivity anisotropy (quantified by RRMSE), while the vertical axis denotes structural complexity measured by the Shannon entropy of hybridization ratios ($sp$/$sp^2$/$sp^3$). The data points are color-coded by the average room-temperature lattice thermal conductivity ($\kappa_L$), with marker sizes proportional to mass density. 
\label{fig9}}
\end{figure}


To contextualize our findings, Fig.~\ref{fig9} positions the discovered allotropes relative to established carbon phases in a multi-dimensional property landscape. This mapping underscores the ability of the AI workflow to access distinct structural topologies and physical extremes. The horizontal axis represents $\kappa_L$ anisotropy, quantified by the Relative Root Mean Square Error (RRMSE),
\begin{equation}
\label{eq1}
\text{RRMSE} = \frac{\sqrt{\frac{1}{3}\sum_{i \in \{a,b,c\}} (\kappa_i - \bar{\kappa})^2}}{\bar{\kappa}}
\end{equation}
where $\kappa_i$ and $\bar{\kappa}$ represent the lattice thermal conductivity along the three principal crystallographic axes and average thermal conductivity. The vertical axis measures the structural complexity via the hybridization Shannon entropy ($H_{hyb}$), calculated based on the proportions of carbon atoms in different hybridization states~\cite{Shannon1948mathematical},
\begin{equation}
\label{eq2}
H_{hyb} = - (\sum_{j \in \{sp, sp^2, sp^3\}} p_j \ln p_j)/ \ln 3
\end{equation}
where $p_j$ is the fraction of atoms with hybridization $j$ and $\ln3$ is the normalization processing. Additionally, the color scale reflects the magnitude of $\bar{\kappa}$, and the size of each data point is proportional to the mass density of the carbon material. Our analysis of structural complexity shown in Fig.~\ref{fig9} reveals a compelling theoretical parallel in the recent work~\cite{Kamil2025Bond}, who proposed that heat transport in coordination-disordered solids is fundamentally governed by the bond-network entropy. Despite the rigidity of its $sp^3$ components, the global lattice thermal conductivity is intrinsically limited by this structural disorder, validating the role of hybridization entropy as a key descriptor for predictive thermal design.


The mapping reveals that unlike diamond and graphene, which reside in low-entropy regions, our generative candidates occupy the high-entropy, high-complexity domain. This complexity enables access to exotic physical properties. A prime example is the $sp$-$sp^2$-$sp^3$ hybrid C$_{12}$, which possesses the highest Shannon entropy and displays anomalous auxeticity. The coexistence of this mechanical anomaly with the metallic nature shown in Fig.~\ref{fig7} highlights the capacity of mixed-hybridization networks to realize multifunctional materials inaccessible via conventional heuristic searches.

Moreover, yne-diamond C$_{12}$ and yne-hex-diamond C$_{8}$ display ultralow in-plane shear moduli combined with high axial thermal conductivity, positioning them as ideal candidates for flexible thermal interface materials~\cite{chen2022interfacial}. The versatility of our approach is further exemplified by C22\_6, which exhibits giant thermal anisotropy, and C16\_3, which resides in the high-density superhard regime. We have thus successfully accessed diverse property domains in carbon materials, including auxetic conductors, superhard insulators, and anisotropic heat spreaders. Ultimately, this work validates the transformative potential of integrating Large Language Models (LLMs) with Machine Learning Potentials to accelerate the inverse design of next-generation functional materials.

As shown in Fig. S4, the cohesive energy of our structures fall near the range of experimentally synthesized metastable carbon allotropes (such as Fullerenes). This energetic proximity suggests they are thermodynamically accessible. More specifically, the cohesive energies of the high-density phases (e.g., C16\_3, C24\_4) are approximately 7.3-7.6 eV/atom, which is comparable to that of the experimentally synthesized Fullerene (C$_{60}$, around 7.6 eV/atom) and significantly higher than amorphous carbons. This places them well within the energetic window of synthesizable metastable phases. On the other hand, the mixed-hybridization phases (yne-diamond C$_{12}$, yne-hex-diamond C$_{8}$, and $sp$-$sp^2$-$sp^3$ hybridized C$_{12}$) exhibit cohesive energies around 7.0 eV/atom. While lower than diamond (around 8.0 eV/atom), this reduction is intrinsic to the incorporation of high-energy acetylenic ($-C\equiv C-$) linkages. Their energetic stability relative to carbyne and graphdiyne. The substantial cohesive energy indicates robust bonding networks resistant to thermal decomposition. We propose that the yne-phases could be realized via bottom-up organic synthesis similar to graphdiyne chemistry~\cite{Li2010Architecture, Matsuoka2017Crystalline, Zhang2017Pseudo}, while the dense superhard phases (C16\_3) are likely accessible via high-pressure confinement of precursors~\cite{Hu2017Compressed}.



In summary, we demonstrate a robust AI-driven workflow for accelerating carbon allotrope discovery. By coupling the generative power of Large Language Models with the speed and accuracy of Machine Learning Potentials, this approach circumvents the efficiency bottlenecks of conventional heuristic searches. Central to this success is our dual-loop active learning strategy, which guarantees high model fidelity across broad hybridization and stress regimes, enabling reliable assessments of dynamical stability and thermal transport. 
%
Applying this pipeline unlocked carbon phases with unique bonding topologies and functionalities. Notably, yne-diamond C$_{12}$ and yne-hex-diamond C$_{8}$ offer high axial thermal conductivity paired with mechanical compliance, targeting flexible electronic applications. Additionally, the identification of the metallic, auxetic $sp$-$sp^2$-$sp^3$ hybrid $C_{12}$ underscores the method's ability to access ``hidden'' regions of the phase diagram. As illustrated in Fig.~\ref{fig9}, our analysis links structural complexity directly to exotic performance. Beyond specific material candidates, this study validates a scalable, AI-driven framework for the inverse design of next-generation functional materials.


\section{Supplementary Material} 
The Supplementary Material contains unstable carbon structures generated by the LLM with corresponding phonon spectra (Figure S1), and the calculated structural, thermal, and mechanical properties of common and designed novel carbon materials (Table S1). It presents the cumulative and spectral lattice thermal conductivity, weighted phase space, and phonon scattering rates of yne-diamond C$_{12}$ at 300 K (Figure S2), the lattice thermal conductivity versus pressure for diamond at 300 K (Figure S3), and the cohesive energy of common carbon materials compared with designed novel carbon materials (Figure S4, Table S2). Additionally, the variation of stress with triaxial equal strain for diamond and graphite (Figure S5), convergence tests of supercell size and ngrid for lattice thermal conductivity calculations (Figure S6), and the renormalized phonon dispersion curves of yne-diamond C$_{12}$, yne-hex-diamond C$_{8}$, and $sp-sp^2-sp^3$ hybridized C$_{12}$ as a function of temperature (Figure S7) are provided. Detailed computational methods regarding structural optimization, data generation, machine learning potential training, and property calculations, along with the optimized crystal structures (POSCAR) for the discussed carbon allotropes are available.

\section{Acknowledgements} 
We acknowledge the support from the National Natural Science Foundation of China (No.52250191). 
We also acknowledge the support by HPC Platform, Xi’an Jiaotong University. The data that support the findings of this study are available from the corresponding authors upon reasonable request.

\section{Author Declarations} 
\subsection{Conflict of interest}
The authors have no conflicts to disclose.





\begin{thebibliography}{59}%
\makeatletter
\providecommand \@ifxundefined [1]{%
 \@ifx{#1\undefined}
}%
\providecommand \@ifnum [1]{%
 \ifnum #1\expandafter \@firstoftwo
 \else \expandafter \@secondoftwo
 \fi
}%
\providecommand \@ifx [1]{%
 \ifx #1\expandafter \@firstoftwo
 \else \expandafter \@secondoftwo
 \fi
}%
\providecommand \natexlab [1]{#1}%
\providecommand \enquote  [1]{``#1''}%
\providecommand \bibnamefont  [1]{#1}%
\providecommand \bibfnamefont [1]{#1}%
\providecommand \citenamefont [1]{#1}%
\providecommand \href@noop [0]{\@secondoftwo}%
\providecommand \href [0]{\begingroup \@sanitize@url \@href}%
\providecommand \@href[1]{\@@startlink{#1}\@@href}%
\providecommand \@@href[1]{\endgroup#1\@@endlink}%
\providecommand \@sanitize@url [0]{\catcode `\\12\catcode `\$12\catcode `\&12\catcode `\#12\catcode `\^12\catcode `\_12\catcode `\%12\relax}%
\providecommand \@@startlink[1]{}%
\providecommand \@@endlink[0]{}%
\providecommand \url  [0]{\begingroup\@sanitize@url \@url }%
\providecommand \@url [1]{\endgroup\@href {#1}{\urlprefix }}%
\providecommand \urlprefix  [0]{URL }%
\providecommand \Eprint [0]{\href }%
\providecommand \doibase [0]{http://dx.doi.org/}%
\providecommand \selectlanguage [0]{\@gobble}%
\providecommand \bibinfo  [0]{\@secondoftwo}%
\providecommand \bibfield  [0]{\@secondoftwo}%
\providecommand \translation [1]{[#1]}%
\providecommand \BibitemOpen [0]{}%
\providecommand \bibitemStop [0]{}%
\providecommand \bibitemNoStop [0]{.\EOS\space}%
\providecommand \EOS [0]{\spacefactor3000\relax}%
\providecommand \BibitemShut  [1]{\csname bibitem#1\endcsname}%
\let\auto@bib@innerbib\@empty
\bibitem [{\citenamefont {Andrievski}(2001)}]{ANDRIEVSKI2001447}%
  \BibitemOpen
  \bibfield  {author} {\bibinfo {author} {\bibfnamefont {R.A.}\ \bibnamefont {Andrievski}},\ }\bibfield  {title} {\enquote {\bibinfo {title} {Superhard materials based on nanostructured high-melting point compounds: achievements and perspectives},}\ }\href {\doibase https://doi.org/10.1016/S0263-4368(01)00023-3} {\bibfield  {journal} {\bibinfo  {journal} {Int. J. Refract. Met. Hard Mat.}\ }\textbf {\bibinfo {volume} {19}},\ \bibinfo {pages} {447--452} (\bibinfo {year} {2001})}\BibitemShut {NoStop}%
\bibitem [{\citenamefont {Iijima}(1991)}]{Iijima1991microtubules}%
  \BibitemOpen
  \bibfield  {author} {\bibinfo {author} {\bibfnamefont {Sumio}\ \bibnamefont {Iijima}},\ }\bibfield  {title} {\enquote {\bibinfo {title} {Helical microtubules of graphitic carbon},}\ }\href {\doibase 10.1038/354056a0} {\bibfield  {journal} {\bibinfo  {journal} {Nature}\ }\textbf {\bibinfo {volume} {354}},\ \bibinfo {pages} {56--58} (\bibinfo {year} {1991})}\BibitemShut {NoStop}%
\bibitem [{\citenamefont {Iijima}\ and\ \citenamefont {Ichihashi}(1993)}]{Iijima1993nanotubes}%
  \BibitemOpen
  \bibfield  {author} {\bibinfo {author} {\bibfnamefont {Sumio}\ \bibnamefont {Iijima}}\ and\ \bibinfo {author} {\bibfnamefont {Toshinari}\ \bibnamefont {Ichihashi}},\ }\bibfield  {title} {\enquote {\bibinfo {title} {Single-shell carbon nanotubes of 1-nm diameter},}\ }\href {\doibase 10.1038/363603a0} {\bibfield  {journal} {\bibinfo  {journal} {Nature}\ }\textbf {\bibinfo {volume} {363}},\ \bibinfo {pages} {603--605} (\bibinfo {year} {1993})}\BibitemShut {NoStop}%
\bibitem [{\citenamefont {Bethune}\ \emph {et~al.}(1993)\citenamefont {Bethune}, \citenamefont {Kiang}, \citenamefont {de~Vries}, \citenamefont {Gorman}, \citenamefont {Savoy}, \citenamefont {Vazquez},\ and\ \citenamefont {Beyers}}]{Bethune1993Cobalt}%
  \BibitemOpen
  \bibfield  {author} {\bibinfo {author} {\bibfnamefont {D.~S.}\ \bibnamefont {Bethune}}, \bibinfo {author} {\bibfnamefont {C.~H.}\ \bibnamefont {Kiang}}, \bibinfo {author} {\bibfnamefont {M.~S.}\ \bibnamefont {de~Vries}}, \bibinfo {author} {\bibfnamefont {G.}~\bibnamefont {Gorman}}, \bibinfo {author} {\bibfnamefont {R.}~\bibnamefont {Savoy}}, \bibinfo {author} {\bibfnamefont {J.}~\bibnamefont {Vazquez}}, \ and\ \bibinfo {author} {\bibfnamefont {R.}~\bibnamefont {Beyers}},\ }\bibfield  {title} {\enquote {\bibinfo {title} {Cobalt-catalysed growth of carbon nanotubes with single-atomic-layer walls},}\ }\href {\doibase 10.1038/363605a0} {\bibfield  {journal} {\bibinfo  {journal} {Nature}\ }\textbf {\bibinfo {volume} {363}},\ \bibinfo {pages} {605--607} (\bibinfo {year} {1993})}\BibitemShut {NoStop}%
\bibitem [{\citenamefont {Balandin}\ \emph {et~al.}(2008)\citenamefont {Balandin}, \citenamefont {Ghosh}, \citenamefont {Bao}, \citenamefont {Calizo}, \citenamefont {Teweldebrhan}, \citenamefont {Miao},\ and\ \citenamefont {Lau}}]{Balandin2008}%
  \BibitemOpen
  \bibfield  {author} {\bibinfo {author} {\bibfnamefont {Alexander~A.}\ \bibnamefont {Balandin}}, \bibinfo {author} {\bibfnamefont {Suchismita}\ \bibnamefont {Ghosh}}, \bibinfo {author} {\bibfnamefont {Wenzhong}\ \bibnamefont {Bao}}, \bibinfo {author} {\bibfnamefont {Irene}\ \bibnamefont {Calizo}}, \bibinfo {author} {\bibfnamefont {Desalegne}\ \bibnamefont {Teweldebrhan}}, \bibinfo {author} {\bibfnamefont {Feng}\ \bibnamefont {Miao}}, \ and\ \bibinfo {author} {\bibfnamefont {Chun~Ning}\ \bibnamefont {Lau}},\ }\bibfield  {title} {\enquote {\bibinfo {title} {Superior thermal conductivity of single-layer graphene},}\ }\href {\doibase 10.1021/nl0731872} {\bibfield  {journal} {\bibinfo  {journal} {Nano Lett.}\ }\textbf {\bibinfo {volume} {8}},\ \bibinfo {pages} {902--907} (\bibinfo {year} {2008})}\BibitemShut {NoStop}%
\bibitem [{\citenamefont {Oganov}\ and\ \citenamefont {Glass}(2006)}]{Oganov2006USPEX}%
  \BibitemOpen
  \bibfield  {author} {\bibinfo {author} {\bibfnamefont {Artem~R.}\ \bibnamefont {Oganov}}\ and\ \bibinfo {author} {\bibfnamefont {Colin~W.}\ \bibnamefont {Glass}},\ }\bibfield  {title} {\enquote {\bibinfo {title} {Crystal structure prediction using ab initio evolutionary techniques: Principles and applications},}\ }\href {\doibase 10.1063/1.2210932} {\bibfield  {journal} {\bibinfo  {journal} {J. Chem. Phys.}\ }\textbf {\bibinfo {volume} {124}},\ \bibinfo {pages} {244704} (\bibinfo {year} {2006})}\BibitemShut {NoStop}%
\bibitem [{\citenamefont {Wang}\ \emph {et~al.}(2010)\citenamefont {Wang}, \citenamefont {Lv}, \citenamefont {Zhu},\ and\ \citenamefont {Ma}}]{Wang2010CALYPSO}%
  \BibitemOpen
  \bibfield  {author} {\bibinfo {author} {\bibfnamefont {Yanchao}\ \bibnamefont {Wang}}, \bibinfo {author} {\bibfnamefont {Jian}\ \bibnamefont {Lv}}, \bibinfo {author} {\bibfnamefont {Li}~\bibnamefont {Zhu}}, \ and\ \bibinfo {author} {\bibfnamefont {Yanming}\ \bibnamefont {Ma}},\ }\bibfield  {title} {\enquote {\bibinfo {title} {Crystal structure prediction via particle-swarm optimization},}\ }\href {\doibase 10.1103/PhysRevB.82.094116} {\bibfield  {journal} {\bibinfo  {journal} {Phys. Rev. B}\ }\textbf {\bibinfo {volume} {82}},\ \bibinfo {pages} {094116} (\bibinfo {year} {2010})}\BibitemShut {NoStop}%
\bibitem [{\citenamefont {Xie}\ and\ \citenamefont {Grossman}(2018)}]{Xie2018}%
  \BibitemOpen
  \bibfield  {author} {\bibinfo {author} {\bibfnamefont {Tian}\ \bibnamefont {Xie}}\ and\ \bibinfo {author} {\bibfnamefont {Jeffrey~C.}\ \bibnamefont {Grossman}},\ }\bibfield  {title} {\enquote {\bibinfo {title} {Crystal graph convolutional neural networks for an accurate and interpretable prediction of material properties},}\ }\href {\doibase 10.1103/PhysRevLett.120.145301} {\bibfield  {journal} {\bibinfo  {journal} {Phys. Rev. Lett.}\ }\textbf {\bibinfo {volume} {120}},\ \bibinfo {pages} {145301} (\bibinfo {year} {2018})}\BibitemShut {NoStop}%
\bibitem [{\citenamefont {Luo}\ \emph {et~al.}(2024)\citenamefont {Luo}, \citenamefont {Wang}, \citenamefont {Gao}, \citenamefont {Lv}, \citenamefont {Wang}, \citenamefont {Chen},\ and\ \citenamefont {Ma}}]{Luo2024Deep}%
  \BibitemOpen
  \bibfield  {author} {\bibinfo {author} {\bibfnamefont {Xiaoshan}\ \bibnamefont {Luo}}, \bibinfo {author} {\bibfnamefont {Zhenyu}\ \bibnamefont {Wang}}, \bibinfo {author} {\bibfnamefont {Pengyue}\ \bibnamefont {Gao}}, \bibinfo {author} {\bibfnamefont {Jian}\ \bibnamefont {Lv}}, \bibinfo {author} {\bibfnamefont {Yanchao}\ \bibnamefont {Wang}}, \bibinfo {author} {\bibfnamefont {Changfeng}\ \bibnamefont {Chen}}, \ and\ \bibinfo {author} {\bibfnamefont {Yanming}\ \bibnamefont {Ma}},\ }\bibfield  {title} {\enquote {\bibinfo {title} {Deep learning generative model for crystal structure prediction},}\ }\href {\doibase 10.1038/s41524-024-01443-y} {\bibfield  {journal} {\bibinfo  {journal} {npj Comput. Mater.}\ }\textbf {\bibinfo {volume} {10}},\ \bibinfo {pages} {254} (\bibinfo {year} {2024})}\BibitemShut {NoStop}%
\bibitem [{\citenamefont {Antunes}\ \emph {et~al.}(2024)\citenamefont {Antunes}, \citenamefont {Butler},\ and\ \citenamefont {Grau-Crespo}}]{antunes2024crystal}%
  \BibitemOpen
  \bibfield  {author} {\bibinfo {author} {\bibfnamefont {Luis~M}\ \bibnamefont {Antunes}}, \bibinfo {author} {\bibfnamefont {Keith~T}\ \bibnamefont {Butler}}, \ and\ \bibinfo {author} {\bibfnamefont {Ricardo}\ \bibnamefont {Grau-Crespo}},\ }\bibfield  {title} {\enquote {\bibinfo {title} {Crystal structure generation with autoregressive large language modeling},}\ }\href {\doibase 10.1038/s41467-024-54639-7} {\bibfield  {journal} {\bibinfo  {journal} {Nat. Commun.}\ }\textbf {\bibinfo {volume} {15}},\ \bibinfo {pages} {10570} (\bibinfo {year} {2024})}\BibitemShut {NoStop}%
\bibitem [{\citenamefont {Song}\ \emph {et~al.}(2025)\citenamefont {Song}, \citenamefont {Fan}, \citenamefont {Lu}, \citenamefont {Ling}, \citenamefont {Zhou},\ and\ \citenamefont {Wang}}]{Song2025Inverse}%
  \BibitemOpen
  \bibfield  {author} {\bibinfo {author} {\bibfnamefont {Zhilong}\ \bibnamefont {Song}}, \bibinfo {author} {\bibfnamefont {Linfeng}\ \bibnamefont {Fan}}, \bibinfo {author} {\bibfnamefont {Shuaihua}\ \bibnamefont {Lu}}, \bibinfo {author} {\bibfnamefont {Chongyi}\ \bibnamefont {Ling}}, \bibinfo {author} {\bibfnamefont {Qionghua}\ \bibnamefont {Zhou}}, \ and\ \bibinfo {author} {\bibfnamefont {Jinlan}\ \bibnamefont {Wang}},\ }\bibfield  {title} {\enquote {\bibinfo {title} {{Inverse design of promising electrocatalysts for CO$_2$ reduction via generative models and bird swarm algorithm}},}\ }\href {\doibase 10.1038/s41467-024-55613-z} {\bibfield  {journal} {\bibinfo  {journal} {Nat. Commun.}\ }\textbf {\bibinfo {volume} {16}},\ \bibinfo {pages} {1053} (\bibinfo {year} {2025})}\BibitemShut {NoStop}%
\bibitem [{\citenamefont {Merchant}\ \emph {et~al.}(2023)\citenamefont {Merchant}, \citenamefont {Batzner}, \citenamefont {Schoenholz}, \citenamefont {Aykol}, \citenamefont {Cheon},\ and\ \citenamefont {Cubuk}}]{Scaling2023Merchant}%
  \BibitemOpen
  \bibfield  {author} {\bibinfo {author} {\bibfnamefont {Amil}\ \bibnamefont {Merchant}}, \bibinfo {author} {\bibfnamefont {Simon}\ \bibnamefont {Batzner}}, \bibinfo {author} {\bibfnamefont {Samuel~S.}\ \bibnamefont {Schoenholz}}, \bibinfo {author} {\bibfnamefont {Muratahan}\ \bibnamefont {Aykol}}, \bibinfo {author} {\bibfnamefont {Gowoon}\ \bibnamefont {Cheon}}, \ and\ \bibinfo {author} {\bibfnamefont {Ekin~Dogus}\ \bibnamefont {Cubuk}},\ }\bibfield  {title} {\enquote {\bibinfo {title} {Scaling deep learning for materials discovery},}\ }\href {\doibase 10.1038/s41586-023-06735-9} {\bibfield  {journal} {\bibinfo  {journal} {Nature}\ }\textbf {\bibinfo {volume} {624}},\ \bibinfo {pages} {80–85} (\bibinfo {year} {2023})}\BibitemShut {NoStop}%
\bibitem [{\citenamefont {Yang}\ \emph {et~al.}(2024)\citenamefont {Yang}, \citenamefont {Hu}, \citenamefont {Zhou}, \citenamefont {Liu}, \citenamefont {Shi}, \citenamefont {Li}, \citenamefont {Li}, \citenamefont {Chen}, \citenamefont {Chen}, \citenamefont {Zeni}, \citenamefont {Horton}, \citenamefont {Pinsler}, \citenamefont {Fowler}, \citenamefont {Zügner}, \citenamefont {Xie}, \citenamefont {Smith}, \citenamefont {Sun}, \citenamefont {Wang}, \citenamefont {Kong}, \citenamefont {Liu}, \citenamefont {Hao},\ and\ \citenamefont {Lu}}]{yang2024mattersim}%
  \BibitemOpen
  \bibfield  {author} {\bibinfo {author} {\bibfnamefont {Han}\ \bibnamefont {Yang}}, \bibinfo {author} {\bibfnamefont {Chenxi}\ \bibnamefont {Hu}}, \bibinfo {author} {\bibfnamefont {Yichi}\ \bibnamefont {Zhou}}, \bibinfo {author} {\bibfnamefont {Xixian}\ \bibnamefont {Liu}}, \bibinfo {author} {\bibfnamefont {Yu}~\bibnamefont {Shi}}, \bibinfo {author} {\bibfnamefont {Jielan}\ \bibnamefont {Li}}, \bibinfo {author} {\bibfnamefont {Guanzhi}\ \bibnamefont {Li}}, \bibinfo {author} {\bibfnamefont {Zekun}\ \bibnamefont {Chen}}, \bibinfo {author} {\bibfnamefont {Shuizhou}\ \bibnamefont {Chen}}, \bibinfo {author} {\bibfnamefont {Claudio}\ \bibnamefont {Zeni}}, \bibinfo {author} {\bibfnamefont {Matthew}\ \bibnamefont {Horton}}, \bibinfo {author} {\bibfnamefont {Robert}\ \bibnamefont {Pinsler}}, \bibinfo {author} {\bibfnamefont {Andrew}\ \bibnamefont {Fowler}}, \bibinfo {author} {\bibfnamefont {Daniel}\ \bibnamefont {Zügner}}, \bibinfo {author} {\bibfnamefont {Tian}\ \bibnamefont {Xie}}, \bibinfo {author} {\bibfnamefont
  {Jake}\ \bibnamefont {Smith}}, \bibinfo {author} {\bibfnamefont {Lixin}\ \bibnamefont {Sun}}, \bibinfo {author} {\bibfnamefont {Qian}\ \bibnamefont {Wang}}, \bibinfo {author} {\bibfnamefont {Lingyu}\ \bibnamefont {Kong}}, \bibinfo {author} {\bibfnamefont {Chang}\ \bibnamefont {Liu}}, \bibinfo {author} {\bibfnamefont {Hongxia}\ \bibnamefont {Hao}}, \ and\ \bibinfo {author} {\bibfnamefont {Ziheng}\ \bibnamefont {Lu}},\ }\bibfield  {title} {\enquote {\bibinfo {title} {Mattersim: A deep learning atomistic model across elements, temperatures and pressures},}\ }\href {https://arxiv.org/abs/2405.04967} {\bibfield  {journal} {\bibinfo  {journal} {arXiv:2405.04967}\ } (\bibinfo {year} {2024})}\BibitemShut {NoStop}%
\bibitem [{\citenamefont {Liu}\ \emph {et~al.}(2025)\citenamefont {Liu}, \citenamefont {Wang}, \citenamefont {Hao}, \citenamefont {Li}, \citenamefont {Sun}, \citenamefont {Lookman}, \citenamefont {Ding},\ and\ \citenamefont {Gao}}]{Liu2025PINK}%
  \BibitemOpen
  \bibfield  {author} {\bibinfo {author} {\bibfnamefont {Yujie}\ \bibnamefont {Liu}}, \bibinfo {author} {\bibfnamefont {Xiaoying}\ \bibnamefont {Wang}}, \bibinfo {author} {\bibfnamefont {Yuzhou}\ \bibnamefont {Hao}}, \bibinfo {author} {\bibfnamefont {Xuejie}\ \bibnamefont {Li}}, \bibinfo {author} {\bibfnamefont {Jun}\ \bibnamefont {Sun}}, \bibinfo {author} {\bibfnamefont {Turab}\ \bibnamefont {Lookman}}, \bibinfo {author} {\bibfnamefont {Xiangdong}\ \bibnamefont {Ding}}, \ and\ \bibinfo {author} {\bibfnamefont {Zhibin}\ \bibnamefont {Gao}},\ }\bibfield  {title} {\enquote {\bibinfo {title} {Pink: physical-informed machine learning for lattice thermal conductivity},}\ }\href {\doibase 10.20517/jmi.2024.86} {\bibfield  {journal} {\bibinfo  {journal} {J. Mater. Inform.}\ }\textbf {\bibinfo {volume} {5}},\ \bibinfo {pages} {12} (\bibinfo {year} {2025})}\BibitemShut {NoStop}%
\bibitem [{\citenamefont {Togo}\ and\ \citenamefont {Tanaka}(2015)}]{TOGO2015First}%
  \BibitemOpen
  \bibfield  {author} {\bibinfo {author} {\bibfnamefont {Atsushi}\ \bibnamefont {Togo}}\ and\ \bibinfo {author} {\bibfnamefont {Isao}\ \bibnamefont {Tanaka}},\ }\bibfield  {title} {\enquote {\bibinfo {title} {First principles phonon calculations in materials science},}\ }\href {\doibase https://doi.org/10.1016/j.scriptamat.2015.07.021} {\bibfield  {journal} {\bibinfo  {journal} {Scr. Mater.}\ }\textbf {\bibinfo {volume} {108}},\ \bibinfo {pages} {1--5} (\bibinfo {year} {2015})}\BibitemShut {NoStop}%
\bibitem [{\citenamefont {Xu}\ \emph {et~al.}(2025)\citenamefont {Xu}, \citenamefont {Bu}, \citenamefont {Pan}, \citenamefont {Lindgren}, \citenamefont {Wu}, \citenamefont {Wang}, \citenamefont {Liu}, \citenamefont {Song}, \citenamefont {Xu}, \citenamefont {Li}, \citenamefont {Hainer}, \citenamefont {Svensson}, \citenamefont {Wiktor}, \citenamefont {Zhao}, \citenamefont {Huang}, \citenamefont {Qian}, \citenamefont {Zhang}, \citenamefont {Zeng}, \citenamefont {Zhang}, \citenamefont {Tang}, \citenamefont {Xiao}, \citenamefont {Yan}, \citenamefont {Shi}, \citenamefont {Liang}, \citenamefont {Wang}, \citenamefont {Liang}, \citenamefont {Cao}, \citenamefont {Wang}, \citenamefont {Ying}, \citenamefont {Xu}, \citenamefont {Chen}, \citenamefont {Zhang}, \citenamefont {Chen}, \citenamefont {Wu}, \citenamefont {Jiang}, \citenamefont {Berger}, \citenamefont {Li}, \citenamefont {Chen}, \citenamefont {Gabourie}, \citenamefont {Dong}, \citenamefont {Xiong}, \citenamefont {Wei}, \citenamefont {Chen}, \citenamefont {Xu},
  \citenamefont {Ding}, \citenamefont {Sun}, \citenamefont {Ala-Nissila}, \citenamefont {Harju}, \citenamefont {Zheng}, \citenamefont {Guan}, \citenamefont {Erhart}, \citenamefont {Sun}, \citenamefont {Ouyang}, \citenamefont {Su},\ and\ \citenamefont {Fan}}]{Xu2025GPUMD40}%
  \BibitemOpen
  \bibfield  {author} {\bibinfo {author} {\bibfnamefont {Ke}~\bibnamefont {Xu}}, \bibinfo {author} {\bibfnamefont {Hekai}\ \bibnamefont {Bu}}, \bibinfo {author} {\bibfnamefont {Shuning}\ \bibnamefont {Pan}}, \bibinfo {author} {\bibfnamefont {Eric}\ \bibnamefont {Lindgren}}, \bibinfo {author} {\bibfnamefont {Yongchao}\ \bibnamefont {Wu}}, \bibinfo {author} {\bibfnamefont {Yong}\ \bibnamefont {Wang}}, \bibinfo {author} {\bibfnamefont {Jiahui}\ \bibnamefont {Liu}}, \bibinfo {author} {\bibfnamefont {Keke}\ \bibnamefont {Song}}, \bibinfo {author} {\bibfnamefont {Bin}\ \bibnamefont {Xu}}, \bibinfo {author} {\bibfnamefont {Yifan}\ \bibnamefont {Li}}, \bibinfo {author} {\bibfnamefont {Tobias}\ \bibnamefont {Hainer}}, \bibinfo {author} {\bibfnamefont {Lucas}\ \bibnamefont {Svensson}}, \bibinfo {author} {\bibfnamefont {Julia}\ \bibnamefont {Wiktor}}, \bibinfo {author} {\bibfnamefont {Rui}\ \bibnamefont {Zhao}}, \bibinfo {author} {\bibfnamefont {Hongfu}\ \bibnamefont {Huang}}, \bibinfo {author} {\bibfnamefont {Cheng}\
  \bibnamefont {Qian}}, \bibinfo {author} {\bibfnamefont {Shuo}\ \bibnamefont {Zhang}}, \bibinfo {author} {\bibfnamefont {Zezhu}\ \bibnamefont {Zeng}}, \bibinfo {author} {\bibfnamefont {Bohan}\ \bibnamefont {Zhang}}, \bibinfo {author} {\bibfnamefont {Benrui}\ \bibnamefont {Tang}}, \bibinfo {author} {\bibfnamefont {Yang}\ \bibnamefont {Xiao}}, \bibinfo {author} {\bibfnamefont {Zihan}\ \bibnamefont {Yan}}, \bibinfo {author} {\bibfnamefont {Jiuyang}\ \bibnamefont {Shi}}, \bibinfo {author} {\bibfnamefont {Zhixin}\ \bibnamefont {Liang}}, \bibinfo {author} {\bibfnamefont {Junjie}\ \bibnamefont {Wang}}, \bibinfo {author} {\bibfnamefont {Ting}\ \bibnamefont {Liang}}, \bibinfo {author} {\bibfnamefont {Shuo}\ \bibnamefont {Cao}}, \bibinfo {author} {\bibfnamefont {Yanzhou}\ \bibnamefont {Wang}}, \bibinfo {author} {\bibfnamefont {Penghua}\ \bibnamefont {Ying}}, \bibinfo {author} {\bibfnamefont {Nan}\ \bibnamefont {Xu}}, \bibinfo {author} {\bibfnamefont {Chengbing}\ \bibnamefont {Chen}}, \bibinfo {author} {\bibfnamefont
  {Yuwen}\ \bibnamefont {Zhang}}, \bibinfo {author} {\bibfnamefont {Zherui}\ \bibnamefont {Chen}}, \bibinfo {author} {\bibfnamefont {Xin}\ \bibnamefont {Wu}}, \bibinfo {author} {\bibfnamefont {Wenwu}\ \bibnamefont {Jiang}}, \bibinfo {author} {\bibfnamefont {Esme}\ \bibnamefont {Berger}}, \bibinfo {author} {\bibfnamefont {Yanlong}\ \bibnamefont {Li}}, \bibinfo {author} {\bibfnamefont {Shunda}\ \bibnamefont {Chen}}, \bibinfo {author} {\bibfnamefont {Alexander~J.}\ \bibnamefont {Gabourie}}, \bibinfo {author} {\bibfnamefont {Haikuan}\ \bibnamefont {Dong}}, \bibinfo {author} {\bibfnamefont {Shiyun}\ \bibnamefont {Xiong}}, \bibinfo {author} {\bibfnamefont {Ning}\ \bibnamefont {Wei}}, \bibinfo {author} {\bibfnamefont {Yue}\ \bibnamefont {Chen}}, \bibinfo {author} {\bibfnamefont {Jianbin}\ \bibnamefont {Xu}}, \bibinfo {author} {\bibfnamefont {Feng}\ \bibnamefont {Ding}}, \bibinfo {author} {\bibfnamefont {Zhimei}\ \bibnamefont {Sun}}, \bibinfo {author} {\bibfnamefont {Tapio}\ \bibnamefont {Ala-Nissila}}, \bibinfo
  {author} {\bibfnamefont {Ari}\ \bibnamefont {Harju}}, \bibinfo {author} {\bibfnamefont {Jincheng}\ \bibnamefont {Zheng}}, \bibinfo {author} {\bibfnamefont {Pengfei}\ \bibnamefont {Guan}}, \bibinfo {author} {\bibfnamefont {Paul}\ \bibnamefont {Erhart}}, \bibinfo {author} {\bibfnamefont {Jian}\ \bibnamefont {Sun}}, \bibinfo {author} {\bibfnamefont {Wengen}\ \bibnamefont {Ouyang}}, \bibinfo {author} {\bibfnamefont {Yanjing}\ \bibnamefont {Su}}, \ and\ \bibinfo {author} {\bibfnamefont {Zheyong}\ \bibnamefont {Fan}},\ }\bibfield  {title} {\enquote {\bibinfo {title} {{GPUMD 4.0: A high-performance molecular dynamics package for versatile materials simulations with machine-learned potentials}},}\ }\href {\doibase https://doi.org/10.1002/mgea.70028} {\bibfield  {journal} {\bibinfo  {journal} {Mat. Gen. Eng. Adv.}\ }\textbf {\bibinfo {volume} {3}},\ \bibinfo {pages} {e70028} (\bibinfo {year} {2025})}\BibitemShut {NoStop}%
\bibitem [{\citenamefont {Fan}\ \emph {et~al.}(2022)\citenamefont {Fan}, \citenamefont {Wang}, \citenamefont {Ying}, \citenamefont {Song}, \citenamefont {Wang}, \citenamefont {Wang}, \citenamefont {Zeng}, \citenamefont {Xu}, \citenamefont {Lindgren}, \citenamefont {Rahm}, \citenamefont {Gabourie}, \citenamefont {Liu}, \citenamefont {Dong}, \citenamefont {Wu}, \citenamefont {Chen}, \citenamefont {Zhong}, \citenamefont {Sun}, \citenamefont {Erhart}, \citenamefont {Su},\ and\ \citenamefont {Ala-Nissila}}]{Fan2022GPUMD}%
  \BibitemOpen
  \bibfield  {author} {\bibinfo {author} {\bibfnamefont {Zheyong}\ \bibnamefont {Fan}}, \bibinfo {author} {\bibfnamefont {Yanzhou}\ \bibnamefont {Wang}}, \bibinfo {author} {\bibfnamefont {Penghua}\ \bibnamefont {Ying}}, \bibinfo {author} {\bibfnamefont {Keke}\ \bibnamefont {Song}}, \bibinfo {author} {\bibfnamefont {Junjie}\ \bibnamefont {Wang}}, \bibinfo {author} {\bibfnamefont {Yong}\ \bibnamefont {Wang}}, \bibinfo {author} {\bibfnamefont {Zezhu}\ \bibnamefont {Zeng}}, \bibinfo {author} {\bibfnamefont {Ke}~\bibnamefont {Xu}}, \bibinfo {author} {\bibfnamefont {Eric}\ \bibnamefont {Lindgren}}, \bibinfo {author} {\bibfnamefont {J.~Magnus}\ \bibnamefont {Rahm}}, \bibinfo {author} {\bibfnamefont {Alexander~J.}\ \bibnamefont {Gabourie}}, \bibinfo {author} {\bibfnamefont {Jiahui}\ \bibnamefont {Liu}}, \bibinfo {author} {\bibfnamefont {Haikuan}\ \bibnamefont {Dong}}, \bibinfo {author} {\bibfnamefont {Jianyang}\ \bibnamefont {Wu}}, \bibinfo {author} {\bibfnamefont {Yue}\ \bibnamefont {Chen}}, \bibinfo {author}
  {\bibfnamefont {Zheng}\ \bibnamefont {Zhong}}, \bibinfo {author} {\bibfnamefont {Jian}\ \bibnamefont {Sun}}, \bibinfo {author} {\bibfnamefont {Paul}\ \bibnamefont {Erhart}}, \bibinfo {author} {\bibfnamefont {Yanjing}\ \bibnamefont {Su}}, \ and\ \bibinfo {author} {\bibfnamefont {Tapio}\ \bibnamefont {Ala-Nissila}},\ }\bibfield  {title} {\enquote {\bibinfo {title} {{GPUMD: A package for constructing accurate machine-learned potentials and performing highly efficient atomistic simulations}},}\ }\href {\doibase 10.1063/5.0106617} {\bibfield  {journal} {\bibinfo  {journal} {J. Chem. Phys.}\ }\textbf {\bibinfo {volume} {157}},\ \bibinfo {pages} {114801} (\bibinfo {year} {2022})}\BibitemShut {NoStop}%
\bibitem [{\citenamefont {Thompson}\ \emph {et~al.}(2022)\citenamefont {Thompson}, \citenamefont {Aktulga}, \citenamefont {Berger}, \citenamefont {Bolintineanu}, \citenamefont {Brown}, \citenamefont {Crozier}, \citenamefont {in~'t Veld}, \citenamefont {Kohlmeyer}, \citenamefont {Moore}, \citenamefont {Nguyen}, \citenamefont {Shan}, \citenamefont {Stevens}, \citenamefont {Tranchida}, \citenamefont {Trott},\ and\ \citenamefont {Plimpton}}]{LAMMPS}%
  \BibitemOpen
  \bibfield  {author} {\bibinfo {author} {\bibfnamefont {A.~P.}\ \bibnamefont {Thompson}}, \bibinfo {author} {\bibfnamefont {H.~M.}\ \bibnamefont {Aktulga}}, \bibinfo {author} {\bibfnamefont {R.}~\bibnamefont {Berger}}, \bibinfo {author} {\bibfnamefont {D.~S.}\ \bibnamefont {Bolintineanu}}, \bibinfo {author} {\bibfnamefont {W.~M.}\ \bibnamefont {Brown}}, \bibinfo {author} {\bibfnamefont {P.~S.}\ \bibnamefont {Crozier}}, \bibinfo {author} {\bibfnamefont {P.~J.}\ \bibnamefont {in~'t Veld}}, \bibinfo {author} {\bibfnamefont {A.}~\bibnamefont {Kohlmeyer}}, \bibinfo {author} {\bibfnamefont {S.~G.}\ \bibnamefont {Moore}}, \bibinfo {author} {\bibfnamefont {T.~D.}\ \bibnamefont {Nguyen}}, \bibinfo {author} {\bibfnamefont {R.}~\bibnamefont {Shan}}, \bibinfo {author} {\bibfnamefont {M.~J.}\ \bibnamefont {Stevens}}, \bibinfo {author} {\bibfnamefont {J.}~\bibnamefont {Tranchida}}, \bibinfo {author} {\bibfnamefont {C.}~\bibnamefont {Trott}}, \ and\ \bibinfo {author} {\bibfnamefont {S.~J.}\ \bibnamefont {Plimpton}},\
  }\bibfield  {title} {\enquote {\bibinfo {title} {{LAMMPS} - a flexible simulation tool for particle-based materials modeling at the atomic, meso, and continuum scales},}\ }\href {\doibase 10.1016/j.cpc.2021.108171} {\bibfield  {journal} {\bibinfo  {journal} {Comp. Phys. Comm.}\ }\textbf {\bibinfo {volume} {271}},\ \bibinfo {pages} {108171} (\bibinfo {year} {2022})}\BibitemShut {NoStop}%
\bibitem [{\citenamefont {Wang}\ \emph {et~al.}(2023)\citenamefont {Wang}, \citenamefont {Gao}, \citenamefont {Zhu}, \citenamefont {Ren}, \citenamefont {Hu}, \citenamefont {Sun}, \citenamefont {Ding}, \citenamefont {Xia},\ and\ \citenamefont {Li}}]{wang2023role}%
  \BibitemOpen
  \bibfield  {author} {\bibinfo {author} {\bibfnamefont {Xiaoying}\ \bibnamefont {Wang}}, \bibinfo {author} {\bibfnamefont {Zhibin}\ \bibnamefont {Gao}}, \bibinfo {author} {\bibfnamefont {Guimei}\ \bibnamefont {Zhu}}, \bibinfo {author} {\bibfnamefont {Jie}\ \bibnamefont {Ren}}, \bibinfo {author} {\bibfnamefont {Lei}\ \bibnamefont {Hu}}, \bibinfo {author} {\bibfnamefont {Jun}\ \bibnamefont {Sun}}, \bibinfo {author} {\bibfnamefont {Xiangdong}\ \bibnamefont {Ding}}, \bibinfo {author} {\bibfnamefont {Yi}~\bibnamefont {Xia}}, \ and\ \bibinfo {author} {\bibfnamefont {Baowen}\ \bibnamefont {Li}},\ }\bibfield  {title} {\enquote {\bibinfo {title} {{Role of high-order anharmonicity and off-diagonal terms in thermal conductivity: A case study of multiphase ${\mathrm{CsPbBr}}_{3}$}},}\ }\href {\doibase 10.1103/PhysRevB.107.214308} {\bibfield  {journal} {\bibinfo  {journal} {Phys. Rev. B}\ }\textbf {\bibinfo {volume} {107}},\ \bibinfo {pages} {214308} (\bibinfo {year} {2023})}\BibitemShut {NoStop}%
\bibitem [{\citenamefont {Matar}\ and\ \citenamefont {Solozhenko}(2024)}]{Matar2024Novel}%
  \BibitemOpen
  \bibfield  {author} {\bibinfo {author} {\bibfnamefont {Samir~F.}\ \bibnamefont {Matar}}\ and\ \bibinfo {author} {\bibfnamefont {Vladimir~L.}\ \bibnamefont {Solozhenko}},\ }\bibfield  {title} {\enquote {\bibinfo {title} {{Novel Superhard Tetragonal Hybrid $sp^3/sp^2$ Carbon Allotropes C$_x$ (x = 5, 6, 7): Crystal Chemistry and $Ab~Initio$ Studies}},}\ }\href {\doibase 10.3390/c10030064} {\bibfield  {journal} {\bibinfo  {journal} {C}\ }\textbf {\bibinfo {volume} {10}},\ \bibinfo {pages} {64} (\bibinfo {year} {2024})}\BibitemShut {NoStop}%
\bibitem [{\citenamefont {Matar}(2023)}]{Samir2023ene}%
  \BibitemOpen
  \bibfield  {author} {\bibinfo {author} {\bibfnamefont {Samir~F.}\ \bibnamefont {Matar}},\ }\bibfield  {title} {\enquote {\bibinfo {title} {{ene-C$_7$ ($sp^3$-$sp^2$) and yne-C$_8$ ($sp^3$-$sp^1$): Novel ultra-hard trigonal hybrid diamonds from crystal engineering and first principles}},}\ }\href {\doibase https://doi.org/10.1016/j.cjph.2023.07.009} {\bibfield  {journal} {\bibinfo  {journal} {Chin. J. Phys.}\ }\textbf {\bibinfo {volume} {85}},\ \bibinfo {pages} {592--599} (\bibinfo {year} {2023})}\BibitemShut {NoStop}%
\bibitem [{\citenamefont {Hu}\ \emph {et~al.}(2014)\citenamefont {Hu}, \citenamefont {Huang}, \citenamefont {Zhao}, \citenamefont {Xu}, \citenamefont {Yu},\ and\ \citenamefont {He}}]{Meng2014Superhard}%
  \BibitemOpen
  \bibfield  {author} {\bibinfo {author} {\bibfnamefont {Meng}\ \bibnamefont {Hu}}, \bibinfo {author} {\bibfnamefont {Quan}\ \bibnamefont {Huang}}, \bibinfo {author} {\bibfnamefont {Zhisheng}\ \bibnamefont {Zhao}}, \bibinfo {author} {\bibfnamefont {Bo}~\bibnamefont {Xu}}, \bibinfo {author} {\bibfnamefont {Dongli}\ \bibnamefont {Yu}}, \ and\ \bibinfo {author} {\bibfnamefont {Julong}\ \bibnamefont {He}},\ }\bibfield  {title} {\enquote {\bibinfo {title} {Superhard and high-strength yne-diamond semimetals},}\ }\href {\doibase https://doi.org/10.1016/j.diamond.2014.04.005} {\bibfield  {journal} {\bibinfo  {journal} {Diam. Relat. Mat.}\ }\textbf {\bibinfo {volume} {46}},\ \bibinfo {pages} {15--20} (\bibinfo {year} {2014})}\BibitemShut {NoStop}%
\bibitem [{\citenamefont {Zhang}\ \emph {et~al.}(2020)\citenamefont {Zhang}, \citenamefont {Chai}, \citenamefont {Fan}, \citenamefont {Song},\ and\ \citenamefont {Yang}}]{ZHANG2020109904}%
  \BibitemOpen
  \bibfield  {author} {\bibinfo {author} {\bibfnamefont {Wei}\ \bibnamefont {Zhang}}, \bibinfo {author} {\bibfnamefont {Changchun}\ \bibnamefont {Chai}}, \bibinfo {author} {\bibfnamefont {Qingyang}\ \bibnamefont {Fan}}, \bibinfo {author} {\bibfnamefont {Yanxing}\ \bibnamefont {Song}}, \ and\ \bibinfo {author} {\bibfnamefont {Yintang}\ \bibnamefont {Yang}},\ }\bibfield  {title} {\enquote {\bibinfo {title} {{A novel two-dimensional $sp$-$sp^2$-$sp^3$ hybridized carbon nanostructure with a negative in-plane Poisson ratio and high electron mobility}},}\ }\href {\doibase https://doi.org/10.1016/j.commatsci.2020.109904} {\bibfield  {journal} {\bibinfo  {journal} {Comput. Mater. Sci.}\ }\textbf {\bibinfo {volume} {185}},\ \bibinfo {pages} {109904} (\bibinfo {year} {2020})}\BibitemShut {NoStop}%
\bibitem [{\citenamefont {Gabourie}\ \emph {et~al.}(2021)\citenamefont {Gabourie}, \citenamefont {Fan}, \citenamefont {Ala-Nissila},\ and\ \citenamefont {Pop}}]{Gabourie2021Spectral}%
  \BibitemOpen
  \bibfield  {author} {\bibinfo {author} {\bibfnamefont {Alexander~J.}\ \bibnamefont {Gabourie}}, \bibinfo {author} {\bibfnamefont {Zheyong}\ \bibnamefont {Fan}}, \bibinfo {author} {\bibfnamefont {Tapio}\ \bibnamefont {Ala-Nissila}}, \ and\ \bibinfo {author} {\bibfnamefont {Eric}\ \bibnamefont {Pop}},\ }\bibfield  {title} {\enquote {\bibinfo {title} {Spectral decomposition of thermal conductivity: Comparing velocity decomposition methods in homogeneous molecular dynamics simulations},}\ }\href {\doibase 10.1103/PhysRevB.103.205421} {\bibfield  {journal} {\bibinfo  {journal} {Phys. Rev. B}\ }\textbf {\bibinfo {volume} {103}},\ \bibinfo {pages} {205421} (\bibinfo {year} {2021})}\BibitemShut {NoStop}%
\bibitem [{\citenamefont {Chen}\ \emph {et~al.}(2025)\citenamefont {Chen}, \citenamefont {Li}, \citenamefont {Zhao}, \citenamefont {Liu}, \citenamefont {Fan}, \citenamefont {Tang},\ and\ \citenamefont {Wang}}]{CHEN2025109859}%
  \BibitemOpen
  \bibfield  {author} {\bibinfo {author} {\bibfnamefont {Chengbing}\ \bibnamefont {Chen}}, \bibinfo {author} {\bibfnamefont {Yutong}\ \bibnamefont {Li}}, \bibinfo {author} {\bibfnamefont {Rui}\ \bibnamefont {Zhao}}, \bibinfo {author} {\bibfnamefont {Zhoulin}\ \bibnamefont {Liu}}, \bibinfo {author} {\bibfnamefont {Zheyong}\ \bibnamefont {Fan}}, \bibinfo {author} {\bibfnamefont {Gang}\ \bibnamefont {Tang}}, \ and\ \bibinfo {author} {\bibfnamefont {Zhiyong}\ \bibnamefont {Wang}},\ }\bibfield  {title} {\enquote {\bibinfo {title} {Neptrain and neptrainkit: Automated active learning and visualization toolkit for neuroevolution potentials},}\ }\href {\doibase https://doi.org/10.1016/j.cpc.2025.109859} {\bibfield  {journal} {\bibinfo  {journal} {Comput. Phys. Commun.}\ }\textbf {\bibinfo {volume} {317}},\ \bibinfo {pages} {109859} (\bibinfo {year} {2025})}\BibitemShut {NoStop}%
\bibitem [{\citenamefont {Li}\ \emph {et~al.}(2010)\citenamefont {Li}, \citenamefont {Li}, \citenamefont {Liu}, \citenamefont {Guo}, \citenamefont {Li},\ and\ \citenamefont {Zhu}}]{Li2010Architecture}%
  \BibitemOpen
  \bibfield  {author} {\bibinfo {author} {\bibfnamefont {Guoxing}\ \bibnamefont {Li}}, \bibinfo {author} {\bibfnamefont {Yuliang}\ \bibnamefont {Li}}, \bibinfo {author} {\bibfnamefont {Huibiao}\ \bibnamefont {Liu}}, \bibinfo {author} {\bibfnamefont {Yanbing}\ \bibnamefont {Guo}}, \bibinfo {author} {\bibfnamefont {Yongjun}\ \bibnamefont {Li}}, \ and\ \bibinfo {author} {\bibfnamefont {Daoben}\ \bibnamefont {Zhu}},\ }\bibfield  {title} {\enquote {\bibinfo {title} {Architecture of graphdiyne nanoscale films},}\ }\href {\doibase 10.1039/B922733D} {\bibfield  {journal} {\bibinfo  {journal} {Chem. Commun.}\ }\textbf {\bibinfo {volume} {46}},\ \bibinfo {pages} {3256--3258} (\bibinfo {year} {2010})}\BibitemShut {NoStop}%
\bibitem [{\citenamefont {Long}\ \emph {et~al.}(2011)\citenamefont {Long}, \citenamefont {Tang}, \citenamefont {Wang}, \citenamefont {Li},\ and\ \citenamefont {Shuai}}]{Long2011Electronic}%
  \BibitemOpen
  \bibfield  {author} {\bibinfo {author} {\bibfnamefont {Mengqiu}\ \bibnamefont {Long}}, \bibinfo {author} {\bibfnamefont {Ling}\ \bibnamefont {Tang}}, \bibinfo {author} {\bibfnamefont {Dong}\ \bibnamefont {Wang}}, \bibinfo {author} {\bibfnamefont {Yuliang}\ \bibnamefont {Li}}, \ and\ \bibinfo {author} {\bibfnamefont {Zhigang}\ \bibnamefont {Shuai}},\ }\bibfield  {title} {\enquote {\bibinfo {title} {Electronic structure and carrier mobility in graphdiyne sheet and nanoribbons: Theoretical predictions},}\ }\href {\doibase 10.1021/nn102472s} {\bibfield  {journal} {\bibinfo  {journal} {ACS Nano}\ }\textbf {\bibinfo {volume} {5}},\ \bibinfo {pages} {2593--2600} (\bibinfo {year} {2011})}\BibitemShut {NoStop}%
\bibitem [{\citenamefont {Zhang}\ \emph {et~al.}(2025)\citenamefont {Zhang}, \citenamefont {Zhao}, \citenamefont {Jiang}, \citenamefont {Cheng},\ and\ \citenamefont {Zhang}}]{Zhang2025Thermal}%
  \BibitemOpen
  \bibfield  {author} {\bibinfo {author} {\bibfnamefont {Jian}\ \bibnamefont {Zhang}}, \bibinfo {author} {\bibfnamefont {Zhuo}\ \bibnamefont {Zhao}}, \bibinfo {author} {\bibfnamefont {Miao}\ \bibnamefont {Jiang}}, \bibinfo {author} {\bibfnamefont {Yuan}\ \bibnamefont {Cheng}}, \ and\ \bibinfo {author} {\bibfnamefont {Gang}\ \bibnamefont {Zhang}},\ }\bibfield  {title} {\enquote {\bibinfo {title} {Thermal properties of hexagonal diamond: Machine learning potential and molecular dynamics study},}\ }\href {\doibase 10.1103/z9gl-yb41} {\bibfield  {journal} {\bibinfo  {journal} {Phys. Rev. Mater.}\ }\textbf {\bibinfo {volume} {9}},\ \bibinfo {pages} {094603} (\bibinfo {year} {2025})}\BibitemShut {NoStop}%
\bibitem [{\citenamefont {Yang}\ \emph {et~al.}(2025)\citenamefont {Yang}, \citenamefont {Lau}, \citenamefont {Zeng}, \citenamefont {Zhang}, \citenamefont {Tang}, \citenamefont {Yan}, \citenamefont {Niu}, \citenamefont {Gou}, \citenamefont {Yang}, \citenamefont {Yang}, \citenamefont {Luo},\ and\ \citenamefont {Mao}}]{Yang2025Synthesis}%
  \BibitemOpen
  \bibfield  {author} {\bibinfo {author} {\bibfnamefont {Liuxiang}\ \bibnamefont {Yang}}, \bibinfo {author} {\bibfnamefont {Kah~Chun}\ \bibnamefont {Lau}}, \bibinfo {author} {\bibfnamefont {Zhidan}\ \bibnamefont {Zeng}}, \bibinfo {author} {\bibfnamefont {Dongzhou}\ \bibnamefont {Zhang}}, \bibinfo {author} {\bibfnamefont {Hu}~\bibnamefont {Tang}}, \bibinfo {author} {\bibfnamefont {Bingmin}\ \bibnamefont {Yan}}, \bibinfo {author} {\bibfnamefont {Guoliang}\ \bibnamefont {Niu}}, \bibinfo {author} {\bibfnamefont {Huiyang}\ \bibnamefont {Gou}}, \bibinfo {author} {\bibfnamefont {Yanping}\ \bibnamefont {Yang}}, \bibinfo {author} {\bibfnamefont {Wenge}\ \bibnamefont {Yang}}, \bibinfo {author} {\bibfnamefont {Duan}\ \bibnamefont {Luo}}, \ and\ \bibinfo {author} {\bibfnamefont {Ho-kwang}\ \bibnamefont {Mao}},\ }\bibfield  {title} {\enquote {\bibinfo {title} {Synthesis of bulk hexagonal diamond},}\ }\href {\doibase 10.1038/s41586-025-09343-x} {\bibfield  {journal} {\bibinfo  {journal} {Nature}\ }\textbf {\bibinfo
  {volume} {644}},\ \bibinfo {pages} {370--375} (\bibinfo {year} {2025})}\BibitemShut {NoStop}%
\bibitem [{\citenamefont {Ge}\ \emph {et~al.}(2025)\citenamefont {Ge}, \citenamefont {Huang}, \citenamefont {Yuan},\ and\ \citenamefont {Zhang}}]{Ge2025Crystalline}%
  \BibitemOpen
  \bibfield  {author} {\bibinfo {author} {\bibfnamefont {Yanqing}\ \bibnamefont {Ge}}, \bibinfo {author} {\bibfnamefont {Shaofeng}\ \bibnamefont {Huang}}, \bibinfo {author} {\bibfnamefont {Zhehao}\ \bibnamefont {Yuan}}, \ and\ \bibinfo {author} {\bibfnamefont {Wei}\ \bibnamefont {Zhang}},\ }\bibfield  {title} {\enquote {\bibinfo {title} {Crystalline porous frameworks via hierarchical dynamic covalent assembly},}\ }\href {\doibase 10.1021/acs.accounts.5c00393} {\bibfield  {journal} {\bibinfo  {journal} {Acc. Chem. Res.}\ }\textbf {\bibinfo {volume} {58}},\ \bibinfo {pages} {2970--2984} (\bibinfo {year} {2025})}\BibitemShut {NoStop}%
\bibitem [{\citenamefont {Zhang}\ \emph {et~al.}(2017{\natexlab{a}})\citenamefont {Zhang}, \citenamefont {Fan}, \citenamefont {Zhang},\ and\ \citenamefont {Fan}}]{Zhang2017Hydrothermal}%
  \BibitemOpen
  \bibfield  {author} {\bibinfo {author} {\bibfnamefont {Wenxia}\ \bibnamefont {Zhang}}, \bibinfo {author} {\bibfnamefont {Baolu}\ \bibnamefont {Fan}}, \bibinfo {author} {\bibfnamefont {Yumeng}\ \bibnamefont {Zhang}}, \ and\ \bibinfo {author} {\bibfnamefont {Jiyang}\ \bibnamefont {Fan}},\ }\bibfield  {title} {\enquote {\bibinfo {title} {{Hydrothermal synthesis of well crystallized C$_8$ and diamond nanocrystals and pH-controlled C$_8$ diamond phase transition}},}\ }\href {\doibase 10.1039/C6CE02635D} {\bibfield  {journal} {\bibinfo  {journal} {CrystEngComm}\ }\textbf {\bibinfo {volume} {19}},\ \bibinfo {pages} {1248--1252} (\bibinfo {year} {2017}{\natexlab{a}})}\BibitemShut {NoStop}%
\bibitem [{\citenamefont {Liu}\ \emph {et~al.}(2008)\citenamefont {Liu}, \citenamefont {Cui},\ and\ \citenamefont {Yang}}]{Liu2008Synthesis}%
  \BibitemOpen
  \bibfield  {author} {\bibinfo {author} {\bibfnamefont {P.}~\bibnamefont {Liu}}, \bibinfo {author} {\bibfnamefont {H.}~\bibnamefont {Cui}}, \ and\ \bibinfo {author} {\bibfnamefont {G.~W.}\ \bibnamefont {Yang}},\ }\bibfield  {title} {\enquote {\bibinfo {title} {Synthesis of body-centered cubic carbon nanocrystals},}\ }\href {\doibase 10.1021/cg7006777} {\bibfield  {journal} {\bibinfo  {journal} {Cryst. Growth Des.}\ }\textbf {\bibinfo {volume} {8}},\ \bibinfo {pages} {581--586} (\bibinfo {year} {2008})}\BibitemShut {NoStop}%
\bibitem [{\citenamefont {Johnston}\ and\ \citenamefont {Hoffmann}(1989)}]{Johnston1989Superdense}%
  \BibitemOpen
  \bibfield  {author} {\bibinfo {author} {\bibfnamefont {Roy~L.}\ \bibnamefont {Johnston}}\ and\ \bibinfo {author} {\bibfnamefont {Roald}\ \bibnamefont {Hoffmann}},\ }\bibfield  {title} {\enquote {\bibinfo {title} {{Superdense carbon, C$_8$: supercubane or analog of $\gamma$-Si?}}}\ }\href {\doibase 10.1021/ja00185a004} {\bibfield  {journal} {\bibinfo  {journal} {J. Am. Chem. Soc.}\ }\textbf {\bibinfo {volume} {111}},\ \bibinfo {pages} {810--819} (\bibinfo {year} {1989})}\BibitemShut {NoStop}%
\bibitem [{\citenamefont {Dong}\ \emph {et~al.}(2023)\citenamefont {Dong}, \citenamefont {Cao}, \citenamefont {Ying}, \citenamefont {Fan}, \citenamefont {Qian},\ and\ \citenamefont {Su}}]{DONG2023123943}%
  \BibitemOpen
  \bibfield  {author} {\bibinfo {author} {\bibfnamefont {Haikuan}\ \bibnamefont {Dong}}, \bibinfo {author} {\bibfnamefont {Chenyang}\ \bibnamefont {Cao}}, \bibinfo {author} {\bibfnamefont {Penghua}\ \bibnamefont {Ying}}, \bibinfo {author} {\bibfnamefont {Zheyong}\ \bibnamefont {Fan}}, \bibinfo {author} {\bibfnamefont {Ping}\ \bibnamefont {Qian}}, \ and\ \bibinfo {author} {\bibfnamefont {Yanjing}\ \bibnamefont {Su}},\ }\bibfield  {title} {\enquote {\bibinfo {title} {Anisotropic and high thermal conductivity in monolayer quasi-hexagonal fullerene: A comparative study against bulk phase fullerene},}\ }\href {\doibase https://doi.org/10.1016/j.ijheatmasstransfer.2023.123943} {\bibfield  {journal} {\bibinfo  {journal} {International Journal of Heat and Mass Transfer}\ }\textbf {\bibinfo {volume} {206}},\ \bibinfo {pages} {123943} (\bibinfo {year} {2023})}\BibitemShut {NoStop}%
\bibitem [{\citenamefont {Peng}\ and\ \citenamefont {Pizzochero}(2025)}]{Peng2025Monolayer}%
  \BibitemOpen
  \bibfield  {author} {\bibinfo {author} {\bibfnamefont {Bo}~\bibnamefont {Peng}}\ and\ \bibinfo {author} {\bibfnamefont {Michele}\ \bibnamefont {Pizzochero}},\ }\bibfield  {title} {\enquote {\bibinfo {title} {{Monolayer C$_{60}$ networks: a first-principles perspective}},}\ }\href {\doibase 10.1039/D5CC02473K} {\bibfield  {journal} {\bibinfo  {journal} {Chem. Commun.}\ }\textbf {\bibinfo {volume} {61}},\ \bibinfo {pages} {10287--10302} (\bibinfo {year} {2025})}\BibitemShut {NoStop}%
\bibitem [{\citenamefont {Peng}(2022)}]{Peng2022Monolayer}%
  \BibitemOpen
  \bibfield  {author} {\bibinfo {author} {\bibfnamefont {Bo}~\bibnamefont {Peng}},\ }\bibfield  {title} {\enquote {\bibinfo {title} {Monolayer fullerene networks as photocatalysts for overall water splitting},}\ }\href {\doibase 10.1021/jacs.2c08054} {\bibfield  {journal} {\bibinfo  {journal} {J. Am. Chem. Soc.}\ }\textbf {\bibinfo {volume} {144}},\ \bibinfo {pages} {19921--19931} (\bibinfo {year} {2022})}\BibitemShut {NoStop}%
\bibitem [{\citenamefont {Hou}\ \emph {et~al.}(2022)\citenamefont {Hou}, \citenamefont {Cui}, \citenamefont {Guan}, \citenamefont {Wang}, \citenamefont {Li}, \citenamefont {Liu}, \citenamefont {Zhu},\ and\ \citenamefont {Zheng}}]{Hou2022Synthesis}%
  \BibitemOpen
  \bibfield  {author} {\bibinfo {author} {\bibfnamefont {Lingxiang}\ \bibnamefont {Hou}}, \bibinfo {author} {\bibfnamefont {Xueping}\ \bibnamefont {Cui}}, \bibinfo {author} {\bibfnamefont {Bo}~\bibnamefont {Guan}}, \bibinfo {author} {\bibfnamefont {Shaozhi}\ \bibnamefont {Wang}}, \bibinfo {author} {\bibfnamefont {Ruian}\ \bibnamefont {Li}}, \bibinfo {author} {\bibfnamefont {Yunqi}\ \bibnamefont {Liu}}, \bibinfo {author} {\bibfnamefont {Daoben}\ \bibnamefont {Zhu}}, \ and\ \bibinfo {author} {\bibfnamefont {Jian}\ \bibnamefont {Zheng}},\ }\bibfield  {title} {\enquote {\bibinfo {title} {Synthesis of a monolayer fullerene network},}\ }\href {\doibase 10.1038/s41586-022-04771-5} {\bibfield  {journal} {\bibinfo  {journal} {Nature}\ }\textbf {\bibinfo {volume} {606}},\ \bibinfo {pages} {507--510} (\bibinfo {year} {2022})}\BibitemShut {NoStop}%
\bibitem [{\citenamefont {Umemoto}\ \emph {et~al.}(2010)\citenamefont {Umemoto}, \citenamefont {Wentzcovitch}, \citenamefont {Saito},\ and\ \citenamefont {Miyake}}]{Umemoto2010Body}%
  \BibitemOpen
  \bibfield  {author} {\bibinfo {author} {\bibfnamefont {Koichiro}\ \bibnamefont {Umemoto}}, \bibinfo {author} {\bibfnamefont {Renata~M.}\ \bibnamefont {Wentzcovitch}}, \bibinfo {author} {\bibfnamefont {Susumu}\ \bibnamefont {Saito}}, \ and\ \bibinfo {author} {\bibfnamefont {Takashi}\ \bibnamefont {Miyake}},\ }\bibfield  {title} {\enquote {\bibinfo {title} {{Body-Centered Tetragonal C$_{4}$: A Viable $sp^{3}$ Carbon Allotrope}},}\ }\href {\doibase 10.1103/PhysRevLett.104.125504} {\bibfield  {journal} {\bibinfo  {journal} {Phys. Rev. Lett.}\ }\textbf {\bibinfo {volume} {104}},\ \bibinfo {pages} {125504} (\bibinfo {year} {2010})}\BibitemShut {NoStop}%
\bibitem [{\citenamefont {Li}(2025)}]{LI2025116070}%
  \BibitemOpen
  \bibfield  {author} {\bibinfo {author} {\bibfnamefont {Yang}\ \bibnamefont {Li}},\ }\bibfield  {title} {\enquote {\bibinfo {title} {{Body-centered tetragonal C$_4$: A carbon allotrope with real topology and second-order bulk-boundary correspondence}},}\ }\href {\doibase https://doi.org/10.1016/j.physe.2024.116070} {\bibfield  {journal} {\bibinfo  {journal} {E Low dimens. Syst. Nanostruct.}\ }\textbf {\bibinfo {volume} {165}},\ \bibinfo {pages} {116070} (\bibinfo {year} {2025})}\BibitemShut {NoStop}%
\bibitem [{\citenamefont {Chen}\ \emph {et~al.}(2011)\citenamefont {Chen}, \citenamefont {Niu}, \citenamefont {Li},\ and\ \citenamefont {Li}}]{CHEN20111275}%
  \BibitemOpen
  \bibfield  {author} {\bibinfo {author} {\bibfnamefont {Xing-Qiu}\ \bibnamefont {Chen}}, \bibinfo {author} {\bibfnamefont {Haiyang}\ \bibnamefont {Niu}}, \bibinfo {author} {\bibfnamefont {Dianzhong}\ \bibnamefont {Li}}, \ and\ \bibinfo {author} {\bibfnamefont {Yiyi}\ \bibnamefont {Li}},\ }\bibfield  {title} {\enquote {\bibinfo {title} {Modeling hardness of polycrystalline materials and bulk metallic glasses},}\ }\href {\doibase https://doi.org/10.1016/j.intermet.2011.03.026} {\bibfield  {journal} {\bibinfo  {journal} {Intermetallics}\ }\textbf {\bibinfo {volume} {19}},\ \bibinfo {pages} {1275--1281} (\bibinfo {year} {2011})}\BibitemShut {NoStop}%
\bibitem [{\citenamefont {Gao}\ \emph {et~al.}(2003)\citenamefont {Gao}, \citenamefont {He}, \citenamefont {Wu}, \citenamefont {Liu}, \citenamefont {Yu}, \citenamefont {Li}, \citenamefont {Zhang},\ and\ \citenamefont {Tian}}]{Gao2003Hardness}%
  \BibitemOpen
  \bibfield  {author} {\bibinfo {author} {\bibfnamefont {Faming}\ \bibnamefont {Gao}}, \bibinfo {author} {\bibfnamefont {Julong}\ \bibnamefont {He}}, \bibinfo {author} {\bibfnamefont {Erdong}\ \bibnamefont {Wu}}, \bibinfo {author} {\bibfnamefont {Shimin}\ \bibnamefont {Liu}}, \bibinfo {author} {\bibfnamefont {Dongli}\ \bibnamefont {Yu}}, \bibinfo {author} {\bibfnamefont {Dongchun}\ \bibnamefont {Li}}, \bibinfo {author} {\bibfnamefont {Siyuan}\ \bibnamefont {Zhang}}, \ and\ \bibinfo {author} {\bibfnamefont {Yongjun}\ \bibnamefont {Tian}},\ }\bibfield  {title} {\enquote {\bibinfo {title} {Hardness of covalent crystals},}\ }\href {\doibase 10.1103/PhysRevLett.91.015502} {\bibfield  {journal} {\bibinfo  {journal} {Phys. Rev. Lett.}\ }\textbf {\bibinfo {volume} {91}},\ \bibinfo {pages} {015502} (\bibinfo {year} {2003})}\BibitemShut {NoStop}%
\bibitem [{\citenamefont {Gao}\ \emph {et~al.}(2017)\citenamefont {Gao}, \citenamefont {Dong}, \citenamefont {Li},\ and\ \citenamefont {Ren}}]{Gao2017Novel}%
  \BibitemOpen
  \bibfield  {author} {\bibinfo {author} {\bibfnamefont {Zhibin}\ \bibnamefont {Gao}}, \bibinfo {author} {\bibfnamefont {Xiao}\ \bibnamefont {Dong}}, \bibinfo {author} {\bibfnamefont {Nianbei}\ \bibnamefont {Li}}, \ and\ \bibinfo {author} {\bibfnamefont {Jie}\ \bibnamefont {Ren}},\ }\bibfield  {title} {\enquote {\bibinfo {title} {Novel two-dimensional silicon dioxide with in-plane negative poisson’s ratio},}\ }\href {\doibase 10.1021/acs.nanolett.6b03921} {\bibfield  {journal} {\bibinfo  {journal} {Nano Letters}\ }\textbf {\bibinfo {volume} {17}},\ \bibinfo {pages} {772--777} (\bibinfo {year} {2017})}\BibitemShut {NoStop}%
\bibitem [{\citenamefont {Li}\ \emph {et~al.}(2014)\citenamefont {Li}, \citenamefont {Carrete}, \citenamefont {{A. Katcho}},\ and\ \citenamefont {Mingo}}]{WuLI2014ShengBTE}%
  \BibitemOpen
  \bibfield  {author} {\bibinfo {author} {\bibfnamefont {Wu}~\bibnamefont {Li}}, \bibinfo {author} {\bibfnamefont {Jesús}\ \bibnamefont {Carrete}}, \bibinfo {author} {\bibfnamefont {Nebil}\ \bibnamefont {{A. Katcho}}}, \ and\ \bibinfo {author} {\bibfnamefont {Natalio}\ \bibnamefont {Mingo}},\ }\bibfield  {title} {\enquote {\bibinfo {title} {Shengbte: A solver of the boltzmann transport equation for phonons},}\ }\href {\doibase https://doi.org/10.1016/j.cpc.2014.02.015} {\bibfield  {journal} {\bibinfo  {journal} {Comput. Phys. Commun.}\ }\textbf {\bibinfo {volume} {185}},\ \bibinfo {pages} {1747--1758} (\bibinfo {year} {2014})}\BibitemShut {NoStop}%
\bibitem [{\citenamefont {Han}\ \emph {et~al.}(2022)\citenamefont {Han}, \citenamefont {Yang}, \citenamefont {Li}, \citenamefont {Feng},\ and\ \citenamefont {Ruan}}]{Han2022FourPhonon}%
  \BibitemOpen
  \bibfield  {author} {\bibinfo {author} {\bibfnamefont {Zherui}\ \bibnamefont {Han}}, \bibinfo {author} {\bibfnamefont {Xiaolong}\ \bibnamefont {Yang}}, \bibinfo {author} {\bibfnamefont {Wu}~\bibnamefont {Li}}, \bibinfo {author} {\bibfnamefont {Tianli}\ \bibnamefont {Feng}}, \ and\ \bibinfo {author} {\bibfnamefont {Xiulin}\ \bibnamefont {Ruan}},\ }\bibfield  {title} {\enquote {\bibinfo {title} {Fourphonon: An extension module to shengbte for computing four-phonon scattering rates and thermal conductivity},}\ }\href {\doibase 10.1016/j.cpc.2021.108179} {\bibfield  {journal} {\bibinfo  {journal} {Comput. Phys. Commun.}\ }\textbf {\bibinfo {volume} {270}},\ \bibinfo {pages} {108179} (\bibinfo {year} {2022})}\BibitemShut {NoStop}%
\bibitem [{\citenamefont {Simoncelli}\ \emph {et~al.}(2019)\citenamefont {Simoncelli}, \citenamefont {Marzari},\ and\ \citenamefont {Mauri}}]{Simoncelli2019}%
  \BibitemOpen
  \bibfield  {author} {\bibinfo {author} {\bibfnamefont {Michele}\ \bibnamefont {Simoncelli}}, \bibinfo {author} {\bibfnamefont {Nicola}\ \bibnamefont {Marzari}}, \ and\ \bibinfo {author} {\bibfnamefont {Francesco}\ \bibnamefont {Mauri}},\ }\bibfield  {title} {\enquote {\bibinfo {title} {Unified theory of thermal transport in crystals and glasses},}\ }\href {\doibase https://doi.org/10.1038/s41567-019-0520-x} {\bibfield  {journal} {\bibinfo  {journal} {Nat. Phys.}\ }\textbf {\bibinfo {volume} {15}},\ \bibinfo {pages} {809--813} (\bibinfo {year} {2019})}\BibitemShut {NoStop}%
\bibitem [{\citenamefont {Feng}\ \emph {et~al.}(2024{\natexlab{a}})\citenamefont {Feng}, \citenamefont {Wang}, \citenamefont {Zhu}, \citenamefont {He}, \citenamefont {Sun}, \citenamefont {Ding}, \citenamefont {Shiomi}, \citenamefont {Xia}, \citenamefont {Li},\ and\ \citenamefont {Gao}}]{feng2024relation}%
  \BibitemOpen
  \bibfield  {author} {\bibinfo {author} {\bibfnamefont {Minxuan}\ \bibnamefont {Feng}}, \bibinfo {author} {\bibfnamefont {Xiaoying}\ \bibnamefont {Wang}}, \bibinfo {author} {\bibfnamefont {Guimei}\ \bibnamefont {Zhu}}, \bibinfo {author} {\bibfnamefont {Cheng}\ \bibnamefont {He}}, \bibinfo {author} {\bibfnamefont {Jun}\ \bibnamefont {Sun}}, \bibinfo {author} {\bibfnamefont {Xiangdong}\ \bibnamefont {Ding}}, \bibinfo {author} {\bibfnamefont {Junichiro}\ \bibnamefont {Shiomi}}, \bibinfo {author} {\bibfnamefont {Yi}~\bibnamefont {Xia}}, \bibinfo {author} {\bibfnamefont {Baowen}\ \bibnamefont {Li}}, \ and\ \bibinfo {author} {\bibfnamefont {Zhibin}\ \bibnamefont {Gao}},\ }\bibfield  {title} {\enquote {\bibinfo {title} {The relation between the atomic mass ratio and quartic anharmonicity in alkali metal hydrides},}\ }\href {\doibase https://doi.org/10.1016/j.mtphys.2024.101423} {\bibfield  {journal} {\bibinfo  {journal} {Mater. Today Phys.}\ }\textbf {\bibinfo {volume} {44}},\ \bibinfo {pages} {101423} (\bibinfo
  {year} {2024}{\natexlab{a}})}\BibitemShut {NoStop}%
\bibitem [{\citenamefont {Xia}\ \emph {et~al.}(2020)\citenamefont {Xia}, \citenamefont {Hegde}, \citenamefont {Pal}, \citenamefont {Hua}, \citenamefont {Gaines}, \citenamefont {Patel}, \citenamefont {He}, \citenamefont {Aykol},\ and\ \citenamefont {Wolverton}}]{Xia2020Throughput}%
  \BibitemOpen
  \bibfield  {author} {\bibinfo {author} {\bibfnamefont {Yi}~\bibnamefont {Xia}}, \bibinfo {author} {\bibfnamefont {Vinay~I.}\ \bibnamefont {Hegde}}, \bibinfo {author} {\bibfnamefont {Koushik}\ \bibnamefont {Pal}}, \bibinfo {author} {\bibfnamefont {Xia}\ \bibnamefont {Hua}}, \bibinfo {author} {\bibfnamefont {Dale}\ \bibnamefont {Gaines}}, \bibinfo {author} {\bibfnamefont {Shane}\ \bibnamefont {Patel}}, \bibinfo {author} {\bibfnamefont {Jiangang}\ \bibnamefont {He}}, \bibinfo {author} {\bibfnamefont {Muratahan}\ \bibnamefont {Aykol}}, \ and\ \bibinfo {author} {\bibfnamefont {Chris}\ \bibnamefont {Wolverton}},\ }\bibfield  {title} {\enquote {\bibinfo {title} {High-throughput study of lattice thermal conductivity in binary rocksalt and zinc blende compounds including higher-order anharmonicity},}\ }\href {\doibase 10.1103/PhysRevX.10.041029} {\bibfield  {journal} {\bibinfo  {journal} {Phys. Rev. X}\ }\textbf {\bibinfo {volume} {10}},\ \bibinfo {pages} {041029} (\bibinfo {year} {2020})}\BibitemShut {NoStop}%
\bibitem [{\citenamefont {Hao}\ \emph {et~al.}(2024)\citenamefont {Hao}, \citenamefont {Zuo}, \citenamefont {Zheng}, \citenamefont {Hou}, \citenamefont {Gu}, \citenamefont {Wang}, \citenamefont {Li}, \citenamefont {Sun}, \citenamefont {Ding},\ and\ \citenamefont {Gao}}]{Hao2024Machine}%
  \BibitemOpen
  \bibfield  {author} {\bibinfo {author} {\bibfnamefont {Yuzhou}\ \bibnamefont {Hao}}, \bibinfo {author} {\bibfnamefont {Yuting}\ \bibnamefont {Zuo}}, \bibinfo {author} {\bibfnamefont {Jiongzhi}\ \bibnamefont {Zheng}}, \bibinfo {author} {\bibfnamefont {Wenjie}\ \bibnamefont {Hou}}, \bibinfo {author} {\bibfnamefont {Hong}\ \bibnamefont {Gu}}, \bibinfo {author} {\bibfnamefont {Xiaoying}\ \bibnamefont {Wang}}, \bibinfo {author} {\bibfnamefont {Xuejie}\ \bibnamefont {Li}}, \bibinfo {author} {\bibfnamefont {Jun}\ \bibnamefont {Sun}}, \bibinfo {author} {\bibfnamefont {Xiangdong}\ \bibnamefont {Ding}}, \ and\ \bibinfo {author} {\bibfnamefont {Zhibin}\ \bibnamefont {Gao}},\ }\bibfield  {title} {\enquote {\bibinfo {title} {Machine learning for predicting ultralow thermal conductivity and high zt in complex thermoelectric materials},}\ }\href {\doibase 10.1021/acsami.4c09043} {\bibfield  {journal} {\bibinfo  {journal} {ACS Appl. Mater. Interfaces}\ }\textbf {\bibinfo {volume} {16}},\ \bibinfo {pages} {47866--47878}
  (\bibinfo {year} {2024})}\BibitemShut {NoStop}%
\bibitem [{\citenamefont {Wang}\ \emph {et~al.}(2024{\natexlab{a}})\citenamefont {Wang}, \citenamefont {Gao}, \citenamefont {Wang}, \citenamefont {Sun}, \citenamefont {Feng}, \citenamefont {Hao}, \citenamefont {Li}, \citenamefont {Zhao},\ and\ \citenamefont {Ding}}]{Wang2024Anomalous}%
  \BibitemOpen
  \bibfield  {author} {\bibinfo {author} {\bibfnamefont {Yang}\ \bibnamefont {Wang}}, \bibinfo {author} {\bibfnamefont {Zhibin}\ \bibnamefont {Gao}}, \bibinfo {author} {\bibfnamefont {Xiaoying}\ \bibnamefont {Wang}}, \bibinfo {author} {\bibfnamefont {Jinping}\ \bibnamefont {Sun}}, \bibinfo {author} {\bibfnamefont {Minxuan}\ \bibnamefont {Feng}}, \bibinfo {author} {\bibfnamefont {Yuzhou}\ \bibnamefont {Hao}}, \bibinfo {author} {\bibfnamefont {Xuejie}\ \bibnamefont {Li}}, \bibinfo {author} {\bibfnamefont {Yinchang}\ \bibnamefont {Zhao}}, \ and\ \bibinfo {author} {\bibfnamefont {Xiangdong}\ \bibnamefont {Ding}},\ }\bibfield  {title} {\enquote {\bibinfo {title} {{Anomalous thermal conductivity in 2D silica nanocages of immobilizing noble gas atom}},}\ }\href {https://doi.org/10.1063/5.0200462} {\bibfield  {journal} {\bibinfo  {journal} {Appl. Phys. Lett.}\ }\textbf {\bibinfo {volume} {124}},\ \bibinfo {pages} {122205} (\bibinfo {year} {2024}{\natexlab{a}})}\BibitemShut {NoStop}%
\bibitem [{\citenamefont {Wang}\ \emph {et~al.}(2024{\natexlab{b}})\citenamefont {Wang}, \citenamefont {Li}, \citenamefont {Feng}, \citenamefont {Li}, \citenamefont {Hao}, \citenamefont {Shi}, \citenamefont {He}, \citenamefont {Ding},\ and\ \citenamefont {Gao}}]{Wang2024thermoelectricity}%
  \BibitemOpen
  \bibfield  {author} {\bibinfo {author} {\bibfnamefont {Xiaoying}\ \bibnamefont {Wang}}, \bibinfo {author} {\bibfnamefont {Mengyang}\ \bibnamefont {Li}}, \bibinfo {author} {\bibfnamefont {Minxuan}\ \bibnamefont {Feng}}, \bibinfo {author} {\bibfnamefont {Xuejie}\ \bibnamefont {Li}}, \bibinfo {author} {\bibfnamefont {Yuzhou}\ \bibnamefont {Hao}}, \bibinfo {author} {\bibfnamefont {Wen}\ \bibnamefont {Shi}}, \bibinfo {author} {\bibfnamefont {Jiangang}\ \bibnamefont {He}}, \bibinfo {author} {\bibfnamefont {Xiangdong}\ \bibnamefont {Ding}}, \ and\ \bibinfo {author} {\bibfnamefont {Zhibin}\ \bibnamefont {Gao}},\ }\bibfield  {title} {\enquote {\bibinfo {title} {{Bonding hierarchy and coordination interaction leading to high thermoelectricity in wide bandgap ${\mathrm{TlAgI}}_{2}$}},}\ }\href {\doibase 10.1103/PhysRevMaterials.8.094601} {\bibfield  {journal} {\bibinfo  {journal} {Phys. Rev. Mater.}\ }\textbf {\bibinfo {volume} {8}},\ \bibinfo {pages} {094601} (\bibinfo {year} {2024}{\natexlab{b}})}\BibitemShut
  {NoStop}%
\bibitem [{\citenamefont {Wang}\ \emph {et~al.}(2024{\natexlab{c}})\citenamefont {Wang}, \citenamefont {Feng}, \citenamefont {Xia}, \citenamefont {Sun}, \citenamefont {Ding}, \citenamefont {Li},\ and\ \citenamefont {Gao}}]{Wang2024Revisiting}%
  \BibitemOpen
  \bibfield  {author} {\bibinfo {author} {\bibfnamefont {Xiaoying}\ \bibnamefont {Wang}}, \bibinfo {author} {\bibfnamefont {Minxuan}\ \bibnamefont {Feng}}, \bibinfo {author} {\bibfnamefont {Yi}~\bibnamefont {Xia}}, \bibinfo {author} {\bibfnamefont {Jun}\ \bibnamefont {Sun}}, \bibinfo {author} {\bibfnamefont {Xiangdong}\ \bibnamefont {Ding}}, \bibinfo {author} {\bibfnamefont {Baowen}\ \bibnamefont {Li}}, \ and\ \bibinfo {author} {\bibfnamefont {Zhibin}\ \bibnamefont {Gao}},\ }\bibfield  {title} {\enquote {\bibinfo {title} {{Revisiting lattice thermal conductivity of CsCl: The crucial role of quartic anharmonicity}},}\ }\href {https://doi.org/10.1063/5.0201393} {\bibfield  {journal} {\bibinfo  {journal} {Appl. Phys. Lett.}\ }\textbf {\bibinfo {volume} {124}},\ \bibinfo {pages} {172201} (\bibinfo {year} {2024}{\natexlab{c}})}\BibitemShut {NoStop}%
\bibitem [{\citenamefont {Feng}\ \emph {et~al.}(2024{\natexlab{b}})\citenamefont {Feng}, \citenamefont {Wang}, \citenamefont {Zhu}, \citenamefont {He}, \citenamefont {Sun}, \citenamefont {Ding}, \citenamefont {Shiomi}, \citenamefont {Xia}, \citenamefont {Li},\ and\ \citenamefont {Gao}}]{FENG2024anharmonicity}%
  \BibitemOpen
  \bibfield  {author} {\bibinfo {author} {\bibfnamefont {Minxuan}\ \bibnamefont {Feng}}, \bibinfo {author} {\bibfnamefont {Xiaoying}\ \bibnamefont {Wang}}, \bibinfo {author} {\bibfnamefont {Guimei}\ \bibnamefont {Zhu}}, \bibinfo {author} {\bibfnamefont {Cheng}\ \bibnamefont {He}}, \bibinfo {author} {\bibfnamefont {Jun}\ \bibnamefont {Sun}}, \bibinfo {author} {\bibfnamefont {Xiangdong}\ \bibnamefont {Ding}}, \bibinfo {author} {\bibfnamefont {Junichiro}\ \bibnamefont {Shiomi}}, \bibinfo {author} {\bibfnamefont {Yi}~\bibnamefont {Xia}}, \bibinfo {author} {\bibfnamefont {Baowen}\ \bibnamefont {Li}}, \ and\ \bibinfo {author} {\bibfnamefont {Zhibin}\ \bibnamefont {Gao}},\ }\bibfield  {title} {\enquote {\bibinfo {title} {The relation between the atomic mass ratio and quartic anharmonicity in alkali metal hydrides},}\ }\href {\doibase https://doi.org/10.1016/j.mtphys.2024.101423} {\bibfield  {journal} {\bibinfo  {journal} {Mater. Today Phys.}\ }\textbf {\bibinfo {volume} {44}},\ \bibinfo {pages} {101423} (\bibinfo
  {year} {2024}{\natexlab{b}})}\BibitemShut {NoStop}%
\bibitem [{\citenamefont {Hao}\ \emph {et~al.}(2025)\citenamefont {Hao}, \citenamefont {Che}, \citenamefont {Wang}, \citenamefont {Li}, \citenamefont {Lookman}, \citenamefont {Sun}, \citenamefont {Ding},\ and\ \citenamefont {Gao}}]{Hao2025Copper}%
  \BibitemOpen
  \bibfield  {author} {\bibinfo {author} {\bibfnamefont {Yuzhou}\ \bibnamefont {Hao}}, \bibinfo {author} {\bibfnamefont {Junwei}\ \bibnamefont {Che}}, \bibinfo {author} {\bibfnamefont {Xiaoying}\ \bibnamefont {Wang}}, \bibinfo {author} {\bibfnamefont {Xuejie}\ \bibnamefont {Li}}, \bibinfo {author} {\bibfnamefont {Turab}\ \bibnamefont {Lookman}}, \bibinfo {author} {\bibfnamefont {Jun}\ \bibnamefont {Sun}}, \bibinfo {author} {\bibfnamefont {Xiangdong}\ \bibnamefont {Ding}}, \ and\ \bibinfo {author} {\bibfnamefont {Zhibin}\ \bibnamefont {Gao}},\ }\bibfield  {title} {\enquote {\bibinfo {title} {{Copper delocalization leads to ultralow thermal conductivity in chalcohalide ${\mathrm{CuBiSeCl}}_{2}$}},}\ }\href {\doibase 10.1103/PhysRevB.111.195207} {\bibfield  {journal} {\bibinfo  {journal} {Phys. Rev. B}\ }\textbf {\bibinfo {volume} {111}},\ \bibinfo {pages} {195207} (\bibinfo {year} {2025})}\BibitemShut {NoStop}%
\bibitem [{\citenamefont {Shannon}(1948)}]{Shannon1948mathematical}%
  \BibitemOpen
  \bibfield  {author} {\bibinfo {author} {\bibfnamefont {C.~E.}\ \bibnamefont {Shannon}},\ }\bibfield  {title} {\enquote {\bibinfo {title} {A mathematical theory of communication},}\ }\href {\doibase 10.1002/j.1538-7305.1948.tb01338.x} {\bibfield  {journal} {\bibinfo  {journal} {Bell Syst. Tech. J}\ }\textbf {\bibinfo {volume} {27}},\ \bibinfo {pages} {379--423} (\bibinfo {year} {1948})}\BibitemShut {NoStop}%
\bibitem [{\citenamefont {Iwanowski}\ \emph {et~al.}(2025)\citenamefont {Iwanowski}, \citenamefont {Cs\'anyi},\ and\ \citenamefont {Simoncelli}}]{Kamil2025Bond}%
  \BibitemOpen
  \bibfield  {author} {\bibinfo {author} {\bibfnamefont {Kamil}\ \bibnamefont {Iwanowski}}, \bibinfo {author} {\bibfnamefont {G\'abor}\ \bibnamefont {Cs\'anyi}}, \ and\ \bibinfo {author} {\bibfnamefont {Michele}\ \bibnamefont {Simoncelli}},\ }\bibfield  {title} {\enquote {\bibinfo {title} {Bond-network entropy governs heat transport in coordination-disordered solids},}\ }\href {\doibase 10.1103/w4p6-b9mp} {\bibfield  {journal} {\bibinfo  {journal} {Phys. Rev. X}\ }\textbf {\bibinfo {volume} {15}},\ \bibinfo {pages} {041041} (\bibinfo {year} {2025})}\BibitemShut {NoStop}%
\bibitem [{\citenamefont {Chen}\ \emph {et~al.}(2022)\citenamefont {Chen}, \citenamefont {Xu}, \citenamefont {Zhou},\ and\ \citenamefont {Li}}]{chen2022interfacial}%
  \BibitemOpen
  \bibfield  {author} {\bibinfo {author} {\bibfnamefont {Jie}\ \bibnamefont {Chen}}, \bibinfo {author} {\bibfnamefont {Xiangfan}\ \bibnamefont {Xu}}, \bibinfo {author} {\bibfnamefont {Jun}\ \bibnamefont {Zhou}}, \ and\ \bibinfo {author} {\bibfnamefont {Baowen}\ \bibnamefont {Li}},\ }\bibfield  {title} {\enquote {\bibinfo {title} {{Interfacial thermal resistance: Past, present, and future}},}\ }\href {https://journals.aps.org/rmp/abstract/10.1103/RevModPhys.94.025002} {\bibfield  {journal} {\bibinfo  {journal} {Rev. Mod. Phys.}\ }\textbf {\bibinfo {volume} {94}},\ \bibinfo {pages} {025002} (\bibinfo {year} {2022})}\BibitemShut {NoStop}%
\bibitem [{\citenamefont {Matsuoka}\ \emph {et~al.}(2017)\citenamefont {Matsuoka}, \citenamefont {Sakamoto}, \citenamefont {Hoshiko}, \citenamefont {Sasaki}, \citenamefont {Masunaga}, \citenamefont {Nagashio},\ and\ \citenamefont {Nishihara}}]{Matsuoka2017Crystalline}%
  \BibitemOpen
  \bibfield  {author} {\bibinfo {author} {\bibfnamefont {Ryota}\ \bibnamefont {Matsuoka}}, \bibinfo {author} {\bibfnamefont {Ryota}\ \bibnamefont {Sakamoto}}, \bibinfo {author} {\bibfnamefont {Ken}\ \bibnamefont {Hoshiko}}, \bibinfo {author} {\bibfnamefont {Sono}\ \bibnamefont {Sasaki}}, \bibinfo {author} {\bibfnamefont {Hiroyasu}\ \bibnamefont {Masunaga}}, \bibinfo {author} {\bibfnamefont {Kosuke}\ \bibnamefont {Nagashio}}, \ and\ \bibinfo {author} {\bibfnamefont {Hiroshi}\ \bibnamefont {Nishihara}},\ }\bibfield  {title} {\enquote {\bibinfo {title} {Crystalline graphdiyne nanosheets produced at a gas/liquid or liquid/liquid interface},}\ }\href {\doibase 10.1021/jacs.6b12776} {\bibfield  {journal} {\bibinfo  {journal} {Journal of the American Chemical Society}\ }\textbf {\bibinfo {volume} {139}},\ \bibinfo {pages} {3145--3152} (\bibinfo {year} {2017})}\BibitemShut {NoStop}%
\bibitem [{\citenamefont {Zhang}\ \emph {et~al.}(2017{\natexlab{b}})\citenamefont {Zhang}, \citenamefont {Wang}, \citenamefont {Zhu}, \citenamefont {Pan}, \citenamefont {Han}, \citenamefont {Li}, \citenamefont {Zhao}, \citenamefont {Ma}, \citenamefont {Wang}, \citenamefont {Su},\ and\ \citenamefont {Niu}}]{Zhang2017Pseudo}%
  \BibitemOpen
  \bibfield  {author} {\bibinfo {author} {\bibfnamefont {Jinying}\ \bibnamefont {Zhang}}, \bibinfo {author} {\bibfnamefont {Rui}\ \bibnamefont {Wang}}, \bibinfo {author} {\bibfnamefont {Xi}~\bibnamefont {Zhu}}, \bibinfo {author} {\bibfnamefont {Aifei}\ \bibnamefont {Pan}}, \bibinfo {author} {\bibfnamefont {Chenxiao}\ \bibnamefont {Han}}, \bibinfo {author} {\bibfnamefont {Xin}\ \bibnamefont {Li}}, \bibinfo {author} {\bibfnamefont {Dan}\ \bibnamefont {Zhao}}, \bibinfo {author} {\bibfnamefont {Chuansheng}\ \bibnamefont {Ma}}, \bibinfo {author} {\bibfnamefont {Wenjun}\ \bibnamefont {Wang}}, \bibinfo {author} {\bibfnamefont {Haibin}\ \bibnamefont {Su}}, \ and\ \bibinfo {author} {\bibfnamefont {Chunming}\ \bibnamefont {Niu}},\ }\bibfield  {title} {\enquote {\bibinfo {title} {Pseudo-topotactic conversion of carbon nanotubes to t-carbon nanowires under picosecond laser irradiation in methanol},}\ }\href {\doibase 10.1038/s41467-017-00817-9} {\bibfield  {journal} {\bibinfo  {journal} {Nat. Commun.}\ }\textbf {\bibinfo
  {volume} {8}},\ \bibinfo {pages} {683} (\bibinfo {year} {2017}{\natexlab{b}})}\BibitemShut {NoStop}%
\bibitem [{\citenamefont {Hu}\ \emph {et~al.}(2017)\citenamefont {Hu}, \citenamefont {He}, \citenamefont {Zhao}, \citenamefont {Strobel}, \citenamefont {Hu}, \citenamefont {Yu}, \citenamefont {Sun}, \citenamefont {Liu}, \citenamefont {Li}, \citenamefont {Ma}, \citenamefont {Kono}, \citenamefont {Shu}, \citenamefont {kwang Mao}, \citenamefont {Fei}, \citenamefont {Shen}, \citenamefont {Wang}, \citenamefont {Juhl}, \citenamefont {Huang}, \citenamefont {Liu}, \citenamefont {Xu},\ and\ \citenamefont {Tian}}]{Hu2017Compressed}%
  \BibitemOpen
  \bibfield  {author} {\bibinfo {author} {\bibfnamefont {Meng}\ \bibnamefont {Hu}}, \bibinfo {author} {\bibfnamefont {Julong}\ \bibnamefont {He}}, \bibinfo {author} {\bibfnamefont {Zhisheng}\ \bibnamefont {Zhao}}, \bibinfo {author} {\bibfnamefont {Timothy~A.}\ \bibnamefont {Strobel}}, \bibinfo {author} {\bibfnamefont {Wentao}\ \bibnamefont {Hu}}, \bibinfo {author} {\bibfnamefont {Dongli}\ \bibnamefont {Yu}}, \bibinfo {author} {\bibfnamefont {Hao}\ \bibnamefont {Sun}}, \bibinfo {author} {\bibfnamefont {Lingyu}\ \bibnamefont {Liu}}, \bibinfo {author} {\bibfnamefont {Zihe}\ \bibnamefont {Li}}, \bibinfo {author} {\bibfnamefont {Mengdong}\ \bibnamefont {Ma}}, \bibinfo {author} {\bibfnamefont {Yoshio}\ \bibnamefont {Kono}}, \bibinfo {author} {\bibfnamefont {Jinfu}\ \bibnamefont {Shu}}, \bibinfo {author} {\bibfnamefont {Ho}~\bibnamefont {kwang Mao}}, \bibinfo {author} {\bibfnamefont {Yingwei}\ \bibnamefont {Fei}}, \bibinfo {author} {\bibfnamefont {Guoyin}\ \bibnamefont {Shen}}, \bibinfo {author} {\bibfnamefont
  {Yanbin}\ \bibnamefont {Wang}}, \bibinfo {author} {\bibfnamefont {Stephen~J.}\ \bibnamefont {Juhl}}, \bibinfo {author} {\bibfnamefont {Jian~Yu}\ \bibnamefont {Huang}}, \bibinfo {author} {\bibfnamefont {Zhongyuan}\ \bibnamefont {Liu}}, \bibinfo {author} {\bibfnamefont {Bo}~\bibnamefont {Xu}}, \ and\ \bibinfo {author} {\bibfnamefont {Yongjun}\ \bibnamefont {Tian}},\ }\bibfield  {title} {\enquote {\bibinfo {title} {Compressed glassy carbon: An ultrastrong and elastic interpenetrating graphene network},}\ }\href {\doibase 10.1126/sciadv.1603213} {\bibfield  {journal} {\bibinfo  {journal} {Sci. Adv.}\ }\textbf {\bibinfo {volume} {3}},\ \bibinfo {pages} {e1603213} (\bibinfo {year} {2017})}\BibitemShut {NoStop}%
\end{thebibliography}
%

\end{document}